\documentclass[a4paper,11pt]{article}
\usepackage{jheppub} % for details on the use of the package, please see the JINST-author-manual
\usepackage{lineno}
\usepackage{cancel}
\usepackage{graphicx}% Include figure files
\usepackage{dcolumn}% Align table columns on decimal point
\usepackage{bm}% bold math
\usepackage[usenames,dvipsnames]{color}
\usepackage{soul}
\usepackage{subcaption}
\usepackage{enumitem}
\usepackage{xspace}
\usepackage{booktabs}
\usepackage{lipsum}
\usepackage[per-mode=power]{siunitx} % Add support for units
\usepackage[ISO]{diffcoeff}
\usepackage[capitalise]{cleveref}
\usepackage{subcaption}
\usepackage{ragged2e}
\usepackage{mathrsfs}
\usepackage{cleveref}
\usepackage{hyperref}
\usepackage[normalem]{ulem}
\usepackage{amsmath}
\usepackage{wrapfig}
\usepackage{booktabs}

\DeclareSIUnit\year{yr} 
\DeclareCaptionJustification{justified}{\justifying}
\newcommand{\bPsi}{{\bm{\mathsf{\Psi}}}}
\newcommand{\bepsilon}{{\bm{\mathsf{\epsilon}}}}

\newcommand{\schr}{\rm Schr{\"o}dinger\xspace}
\newcommand{\bk}{{\bm k}}

\newcommand\funop[1]{\mathop{{}#1}}
\newcommand{\m}{m} % Math mode

\definecolor{rp}{cmyk}{0.2, 1, 0.6, 0}
\definecolor{rp}{cmyk}{0.2, 1, 0.6, 0}
\definecolor{green2}{cmyk}{0.27, 0, 1, 0.52}

\usepackage{color}
\hypersetup{
    colorlinks=true,       % false: boxed links; true: colored links
    linkcolor=green2,          % color of internal links
    citecolor=green2,        % color of links to bibliography
    filecolor=magenta,      % color of file links
    urlcolor=green2           % color of external links
}
\newcommand{\bvec}[1]{\bm{#1}}

\title{Vector Wave Dark Matter and \\ Terrestrial Quantum Sensors} 
\author[a*]{Dorian W.~P.~Amaral}
\emailAdd{dorian.amaral@rice.edu}
\author[a,b*]{, Mudit Jain}
\emailAdd{mudit.jain@kcl.ac.uk}
\author[a]{\newline Mustafa A. Amin}
\emailAdd{mustafa.a.amin@rice.edu}
\author[a]{, and Christopher Tunnell}
\emailAdd{christopher.tunnell@rice.edu}

\affiliation[a]{Department of Physics and Astronomy, Rice University,
Houston, TX, 77005, U.S.A.}
\affiliation[b]{Theoretical Particle Physics and Cosmology, King’s College London, Strand, London, WC2R 2LS, United Kingdom}
\affiliation[*]
{These authors contributed approximately equally to this work.}

\abstract{(Ultra)light spin-$1$ particles---dark photons---can constitute all of dark matter (DM) and have beyond Standard Model  couplings. This can lead to a coherent, oscillatory signature in terrestrial detectors that depends on the coupling strength. We provide a signal analysis and statistical framework for inferring the properties of such DM by taking into account (i) the stochastic and (ii) the vector nature of the underlying field, along with (iii) the effects due to the Earth's rotation. Owing to equipartition, on time scales shorter than the coherence time the DM field vector typically traces out a fixed ellipse. Taking this ellipse and the rotation of the Earth into account, we highlight a distinctive three-peak signal in Fourier space that can be used to constrain DM coupling strengths. Accounting for all three peaks, we derive latitude-independent constraints on such DM couplings, unlike those stemming from single-peak studies. We apply our framework to the search for ultralight $B - L$ DM using optomechanical sensors, demonstrating the ability to delve into previously unprobed regions of this DM candidate's parameter space.}

\begin{document}

\begin{flushright}
\raggedleft
    \small KCL-PH-TH/2024-09
\end{flushright}

\maketitle
\flushbottom

\newpage

%%%%%%%%%%%%%%%%%%%%%%%%%%%%%%%%%%%%%%%%%%%%%%%%%%%%%%%%%%%%%%%%%%%
\section{Introduction}
\label{sec:intro}
%%%%%%%%%%%%%%%%%%%%%%%%%%%%%%%%%%%%%%%%%%%%%%%%%%%%%%%%%%%%%%%%%%%

Dark matter (DM) dominates the non-relativistic matter content in our cosmos. However, we know exceptionally little about the constituent particles/fields of DM. Apart from the fact that they must interact gravitationally, we do not know their mass, spin, and other potential interactions~\cite{Bertone:2016nfn,Freese:2017idy}. Astrophysical observations allow for a broad range of masses for the dark matter ``particles": $10^{-19}{\rm eV}\lesssim {\rm few} \times M_{\odot}$~\cite{Brandt:2016aco,Dalal:2022rmp}. Theoretical models include particle masses that span this range, with ultralight bosons at the lower end and composite particles/primordial black holes at the upper end~\cite{Hawking:1971ei,Chapline:1975ojl,Hu:2000ke,Frampton:2010sw,Carr:2020gox,Carr:2020xqk}.

Among the variety of possibilities, the case of ultralight, bosonic dark matter is particularly intriguing. These include, for instance, the QCD axion \cite{Weinberg:1977ma,Wilczek:1977pj,DiLuzio:2020wdo}, axion-like particles and other scalars \cite{Arvanitaki:2009fg,Ringwald:2014vqa,Hui:2016ltb,Ferreira:2020fam}, and vector particles \cite{Jaeckel:2012mjv,Fabbrichesi:2020wbt}. A wide-ranging observational and experimental program is currently exploring models that can be tested with contemporary technology. Assuming a local dark matter density $\rho \sim$ GeV cm$^{-3}$ with a typical virial velocity of $v_0 \sim 10^{-3} c$, for particle masses smaller than a few eV, the typical particle number within a de Broglie volume becomes sufficiently large, allowing for a classical field theory description: $N_{\mathrm{dB}} \simeq {\rho}/{\m} \left({h}/{\m v_0}\right)^3  \sim 10^{66} \left({10^{-15}\,{\rm eV}}/{m}\right)^4$.

A widely pursued DM candidate is the (ultra-)light vector dark matter (VDM) particle. Several early-universe production mechanisms exist for such dark matter \cite{Graham:2015rva,Agrawal:2018vin,Co:2018lka,Dror:2018pdh,Bastero-Gil:2018uel,Long:2019lwl,Kolb:2020fwh,Co:2021rhi,Adshead:2023qiw,Cyncynates:2023zwj,Ozsoy:2023gnl} and, recently, numerical simulations of structure formation of light vector dark matter in the nonlinear regime have been carried out (e.g.~\cite{Amin:2022pzv,Gorghetto:2022sue,Jain:2023ojg,Chen:2023bqy}). Our focus here is on the detection prospects of this kind of dark matter. Many dedicated studies have been conducted on the detection or exclusion of vectors \cite{Pierce:2018xmy,Badurina:2019hst,Carney:2019cio,Michimura:2020vxn,Manley:2020mjq,Chen:2021bdr,Caputo:2021eaa,Abe_2021,Nakatsuka:2022gaf} as well as scalars \cite{Foster:2017hbq,Badurina:2019hst,Chen:2021bdr,LIGOScientific:2021ffg,Lisanti:2021vij,Badurina:2021lwr,Badurina:2022ngn,Nakatsuka:2022gaf, Badurina:2023wpk,Miller:2023kkd}. However, for the vector case, the analyses focusing on the Fourier space signal did not fully account for the vector nature and stochastic aspects of the field in their statistical treatment, and nor did they model the effect of the rotation of the Earth \cite{Carney:2019cio,Manley:2020mjq}. In this work, we take these effects into account when studying the sensitivity of mono-directional accelerometers to ultralight vector dark matter.

In our detection scheme, the sensor points in a fixed direction relative to the local tangent plane of the experiment, rotating with the Earth at an angular frequency 
${\omega_{\oplus} \equiv  2\pi / ({\rm 1~sidereal~day}) \approx \SI{7.5e-5}{\hertz}}$. We will show that the signal in Fourier space contains three distinct peaks located at the angular frequencies $m$, $m - \omega_\oplus$, and $ m + \omega_\oplus$, with the last two arising due to the Earth's rotation.\footnote{For the detection scheme and dark matter model we focus on in this work, the signal will manifest as three peaks; however, more peaks can present themselves in general. For example, oscillatory signals at LIGO can lead to five peaks arising in Fourier space \cite{Piccinni:2018akm}.} Since $\omega_{\oplus}$ corresponds to a mass scale of $\sim \SI{5e{-20}}{\electronvolt}$, which is already outside of the allowed mass window, we naturally have $m > \omega_{\oplus}$. To resolve these three distinct peaks, we shall therefore enforce that the observation time always satisfies $T_{\rm obs} > \SI{1}{\day}$. With this basic setup, we will concentrate on the short observation time regime in this paper: experimental expedition timescales which are much smaller than the coherent timescale of the wave DM field, $T_{\rm obs} \ll \tau_{\rm coh} \equiv h /mv_0^2$. 
Thus, our observation time will always lie between
\begin{align}
    1\,{\si{\day}} \lesssim T_{\rm obs} \ll 10\,{\si{\day}}\left(\frac{5 \times 10^{-15}\,{\rm eV}}{m}\right) \,.
\end{align}
This means that, at most, our framework is valid for masses that give a coherence time of at least $\SI{1}{\day}$, corresponding to $m \lesssim 5\times10^{-15}$ eV. For longer expedition times, this upper mass limit decreases: for example, for a runtime of $1\,\mathrm{yr}$, we have that $m \lesssim \SI{e-16}{\electronvolt}$. This gives us the mass range of $\SI{5e-20}{\electronvolt} \lesssim m \lesssim \SI{5e-15}{\electronvolt}$ where our analysis is appropriate\footnote{Later, in~\cref{sec:expand_mass_window}, we shall discuss a plausible detection scheme where the sensor is instead made to rotate manually at higher frequencies, allowing us to probe masses above $\sim 5\times{10^{-15}}$ eV.}.

We limit ourselves to working within the coherent regime for two reasons. Foremost, we wish to highlight how the inherent stochasticity of the full 3-dimensional vector field should be treated when drawing inferences. This randomness is roughly only manifest for observation times shorter than a coherence time since the random variables dictating the behaviour of the field can be treated to be sampled once every coherence patch. For longer observation times, this randomness is averaged over. Secondly, remaining in the coherent regime allows us to treat the signal as only appearing within three bins in Fourier space. For observation times outside of this regime, we would instead begin to resolve the shape of the dark matter halo velocity distribution, complicating our inferencing. Constraining ourselves to this regime thus simplifies the problem, emphasizing the role that the vector nature of this type of dark matter plays when performing inferences. 

Given the recent and substantial research and development efforts in quantum technologies, we are specifically interested in timely studies aimed at understanding the potential of quantum optomechanical sensors in the direct detection of dark matter. Mechanical detectors have a rich history in tests of gravity, including LIGO, and in recent years there has been a surge in efforts to explore their potential in quantum sensing for fundamental physics investigations (see reviews \cite{Ahmed:2018oog,RevModPhys.90.025008, Beckey:2023shi}). We are only beginning to understand the new opportunities for dark matter searches \cite{Carney:2020xol,Higgins:2023gwq,Baker:2023kwz,Kilian:2024fsg,Moore:2020awi,Monteiro:2020wcb} in light of significant advances in quantum readout and control of mechanical sensing devices using optical or microwave light \cite{BLENCOWE2004159,RevModPhys.86.1391,Buchmueller:2022djy}. Accurately modeling the dark matter signal and the associated statistics is crucial for drawing representative inferences, guiding these quantum research and development efforts and aiding in experimental design. This is particularly relevant for large-scale accelerometer projects, such as the one proposed by the Windchime collaboration~\cite{Windchime:2022whs}. Such sensors have demonstrated potential as powerful probes for the wavelike signature produced by ultralight dark matter \cite{Graham:2015ifn}.

In this paper, we devise the analysis strategy for ultralight vector dark matter in the coherent regime to draw more representative exclusion inferences in the future. We begin by laying the theoretical groundwork for this DM paradigm in~\cref{sec:main_calc}. Considering equipartition between the longitudinal and transverse modes of the VDM field, we derive the associated signal in both the time and frequency domains. We do this by taking into account both its stochastic and polarization properties, as well as accounting for the rotation of the Earth. We then perform statistical analyses of the DM signal in the frequency domain in~\cref{sec:stats}. We derive a limit on a generalised parameter that is independent of the vector dark matter model and experimental parameters, which can be recast to concrete choices of them. Unlike other studies that focus solely on a single peak, our findings reveal that the signal power is distributed across three distinct peaks. Accounting for this distribution ensures the retention of constraining power, regardless of the experiment's location on Earth. Finally, in~\cref{sec:application}, we apply our framework to a concrete dark matter model and sensor: $B - L$ dark matter and the canonical optomechanical light cavity. 

We have also included $5$ appendices in this paper. In~\cref{app:long_trans_modes_VDM}, we discuss the stochastic behaviour of the vector field given an unequal distribution of power amongst its longitudinal and transverse modes, as well as their ultimate equipartition (owing to non-linear gravitational dynamics). In~\cref{app:deriv-marginal}, we derive the marginal likelihood for the three-peak signal and also show that the powers in the three peaks are uncorrelated. In~\cref{app:likelihood_gradientscalar}, we discuss the applicability of our results in the context of the gradient of a scalar, deriving limits on the appropriate generalized parameter. In~\cref{app:likelihood_linearpol}, we derive the likelihood in the case that the vector field is `linearly polarized' and also show that the covariance matrix is not diagonal. Finally, in~\cref{app:microscope}, we derive an updated limit on the gauge coupling of a new, long-range $B-L$ coupled fifth force given the latest MICROSCOPE results. Throughout the rest of this paper, we will work in natural units, whereby $\hbar = c = 1$. Moreover, we sometimes quote the DM mass in units of \si{\hertz} where it is more appropriate to treat it as an angular frequency. The conversion from \si{\hertz} to \si{\electronvolt} is given by the relation $m = \hbar \omega / c^2 \approx \SI{4.14e-15}{\electronvolt} [\omega / (2\pi\,\mathrm{Hz})]$.

%%%%%%%%%%%%%%%%%%%%%%%%%%%%%%%%%%%%%%%%%%%%%%%%%%%%%%%%%%%%%%%%%%%
\section{Calculating the Stochastic Wave Vector Dark Matter Signal}
\label{sec:main_calc}
%%%%%%%%%%%%%%%%%%%%%%%%%%%%%%%%%%%%%%%%%%%%%%%%%%%%%%%%%%%%%%%%%%%
\subsection{The Dark Photon Field}
The random vector field $\hat{{\bm A}}$ of mass $m$, at any given location and time, can be decomposed as
\begin{align}
    \hat{{\bm A}}({\bm x},t) = \frac{1}{\sqrt{2m}}e^{-imt}\hat{{\bPsi}}({\bm x},t) + \mathrm{c.c.}\,,
\end{align}
where, in the non-relativistic limit and assuming free field evolution, each Fourier mode of the complex $3$-vector field $\hat{{\bPsi}}({\bm x},t)$ evolves as $\hat{\bPsi}_{\bk}(t) = \hat{\bPsi}_{\bm k}\,e^{-i\frac{k^2}{2m}t}$. We use hatted notation to indicate that a quantity is stochastic.

With non-linear gravitational clustering, we expect an equipartition between longitudinal and transverse polarizations of the plane waves in our vicinity, regardless of whether the early universe production mechanism favors one over the other. In~\cref{app:long_trans_modes_VDM}, we discuss this equipartition in greater detail, providing evidence for it via halo-formation $3$D simulations of the \schr field $\bPsi$.

Given this equipartition, the spectrum
$\langle\hat{\bPsi}^{\dagger}_{\bm k}\cdot\hat{\bPsi}_{\bm p}\rangle = \delta_{\bm k,\,\bm p}\,f_{\bm k}$, where $f_{\bm k}$ is the Milky Way halo distribution function\footnote{The halo function can be given as $f_{\bm k} = [\mathcal{N}(2\pi/(m\sigma V^{1/3}))]^{-3}(\rho V/m)e^{-({\bm k} - \bar{\bm k})^2/(2m^2\sigma^2)}$ where $\mathcal{N}(x) = {\rm ET}(3,\, 0,\, e^{-x^2/2})$ is the elliptic theta function (which goes as $\sqrt{2\pi}/x$ as $x \rightarrow 0$). Here, $\bar{\bm k}/m$ is our velocity w.r.t.~the rest frame of the halo, and $\sigma$ is the velocity dispersion.}
with $V^{-1}\sum_{\bm k}f_{\bm k} = \rho/m$. Here, $V$ is the volume and $\rho$ is the local mass density. To be explicit, we work with a finite volume $V$ so that ${\bm k}$ is discretized. We can define $\hat{\bPsi}_{\bm k}=\sqrt{f_{\bm k}}\,\hat{\bepsilon}_{\bm k}$ such that, for every ${\bm k}$, $\hat{\bepsilon}_{\bm k}$ is a set of $6$ real ($3$ complex) i.i.d.s (independent and identically distributed random variables) with unit norm $\langle\hat{\bepsilon}^{\dagger}_{\bm k}\cdot\hat{\bepsilon}_{\bm p}\rangle = \delta_{\bm k, \bm p}$. In other words, for every ${\bm k}$ there are $5$ real random numbers that are uniformly distributed on a unit $S_5$. With this, a realization of the random vector field is
\begin{align}
    \hat{{\bm A}}({\bm x},t) = \sqrt{\frac{2}{m}}\,\Re\Biggl\{\frac{1}{\sqrt{V}}\sum_{\bm k}\,e^{i{\bm k}\cdot{\bm x}}\,\sqrt{f_{\bm k}}\,\hat{\bepsilon}_{\bm k}\,e^{-i\left(m + \frac{k^2}{2m}\right)t}\Biggr\}\,.
\end{align}
In this paper, we are interested in the short observation time limit, $T_\mathrm{obs} \ll \tau_{\rm coh} = m/k_0^2$ (where $k_0 = mv_0$ denotes the typical wavenumber). In this case, we can neglect the $k^2/2m$ factor from the time-varying sinusoid. Subsequently, we have a summation over many monochromatic waves (all oscillating with frequency $m$), with different amplitudes and phases for different values of ${\bm k}$. Assuming that the halo function is well behaved (meaning any $n^{\rm th}$ moment $\sum_{\bm k} k^n\,f_{\bm k}$ is finite), we can use the central limit theorem to arrive at the following\footnote{We note that, if it were not for equiparition, the statistics for $\bvec{w}$ would follow a more complicated form. We derive this in \cref{app:long_trans_modes_VDM}. In particular, see \cref{eq:cc_corr}.}:
\begin{equation}
\label{eq:A_decompose_main}
    \begin{split}
    &\hat{{\bm A}}({\bm x},t)\Bigr|_{t \ll \tau_{\rm coh}} \approx \sqrt{\frac{2}{m}}\,\Re\Biggl\{e^{-imt}\,\hat{\bm w}({\bm x})\Biggr\}\,,\\
    {\rm where}\qquad &\langle\hat{\bm w}({\bm x})\rangle = 0\quad{\rm and}\quad \langle\hat{w}^{i\,\ast}({\bm x})\hat{w}^{j}({\bm y})\rangle = \frac{\delta^{ij}}{3} \times\frac{1}{V}\sum_{\bm k}e^{i{\bm k}\cdot({\bm x}-{\bm y})}f_{\bm k}\,.
    \end{split}
\end{equation}

For $T_{\rm obs} \ll \tau_{\rm coh}$, we can also safely assume that the distance Earth sweeps during the observation time is negligible compared to the de Broglie length. As such, we shall set ${\bm x} = {\bm y}$ and further set ${\bm x} = 0$ assuming statistical homogeneity of the DM field. Decomposing the complex Gaussian random variables $\hat{\bm w}$ into Euler form, $\hat{w}^j = \hat{\alpha}^j\,e^{-i\hat{\varphi}^j}$, we may write
\begin{align}
\label{eq:Afield_lowT}
    \hat{A}^j(t)\Bigr|_{t \ll \tau_{\rm coh}} \approx \sqrt{\frac{2\rho}{3 m^2}}\,\hat{\alpha}^j\cos\left(mt + \hat{\varphi}^j
    \right)\,.
\end{align}
The three $\hat{\alpha}^j$ are independent Rayleigh distributed random variables, while the three $\varphi^j$ are independent uniformly distributed angles (ranging from $0$ to $2\pi$): 
\begin{equation}
\label{eq:distributions}
    \funop{P}(\alpha^j) = 2\alpha^je^{-{(\alpha^j)}^2} \qquad \text{and} \qquad P(\varphi^j) = \frac{1}{2\pi}\,.
\end{equation}
The three components become statistically independent, mimicking three independent scalars.\footnote{This can be generalized to the many-components case. In particular, for multi-component scalar dark matter (with naturally different masses and also different energy densities for each component), we can simply replace the mass and density by their respective values, $m^j$ and $\rho^j$.} Also note that our result is similar to the case of the gradient of a scalar~\cite{Lisanti:2021vij}, in the sense that there are $6$ independent normal random variables to describe the DM field at a given location and short time scales.

\subsubsection{Equipartion and Ellipses}
\label{subsubsec:equip}

In this subsection, we further justify our assumption of equipartition in the previous section.\\

\begin{figure}[t!]
    \centering
    \includegraphics[width=6 in]{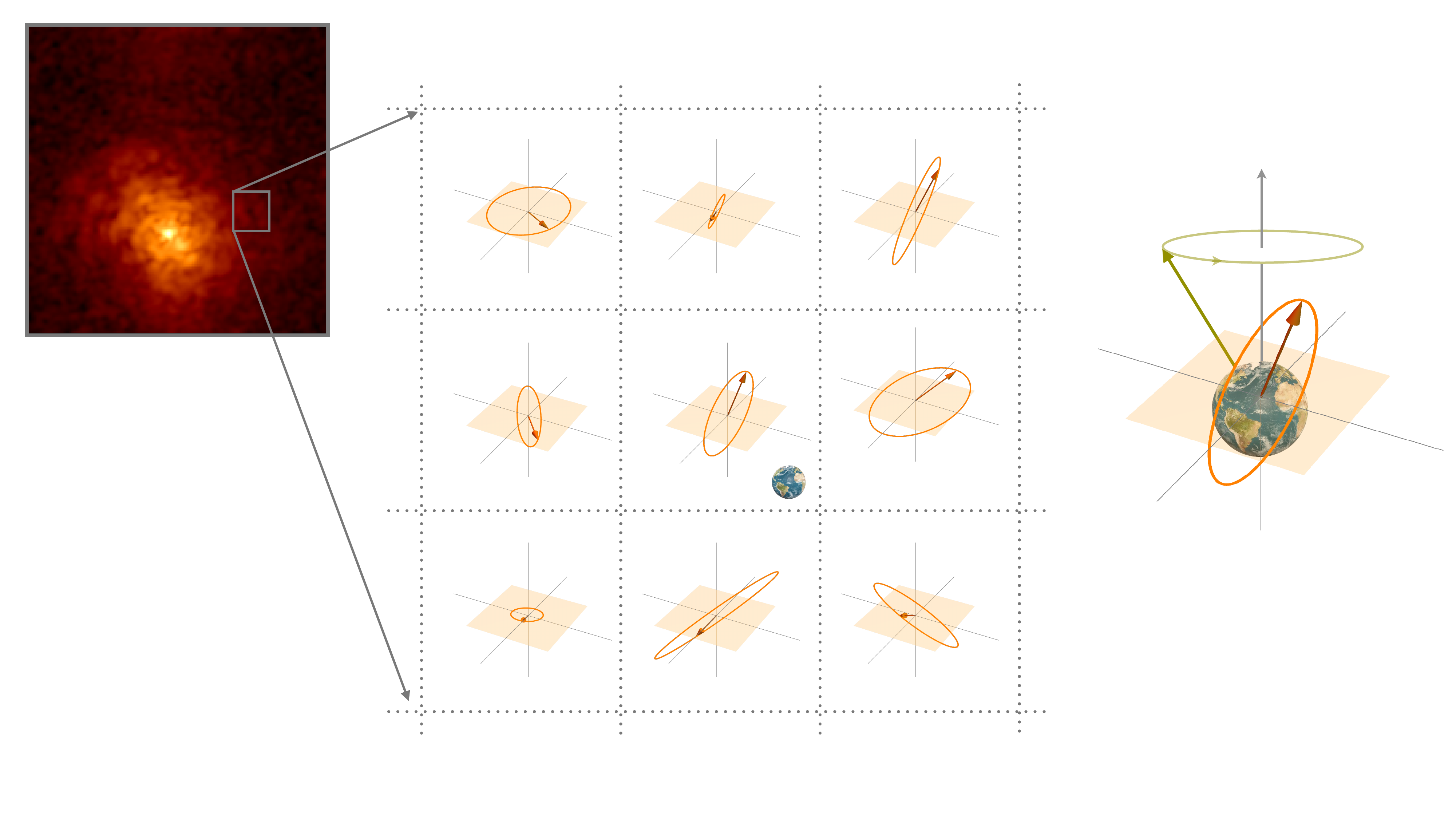}
    \caption{At each spatial point with a coherence region of size $2\pi/mv_0$ (scale of interference ``granules"), the dark matter field $\bm A$ (orange vector) traces out an approximately fixed ellipse (for $2\pi/mc^2=\tau_{\rm Comp}\ll t\ll\tau_{\rm coh}=2\pi/mv_0^2$). These ellipses change their size and orientations on the coherence time scale and smoothly connect with each other from one coherence region to another. For the duration of the measurement, we expect to find ourselves within one such coherence region. The detection signal $\mathcal{S}(t)$ is proportional to the dot product of the dark electric field $\bvec{E} \simeq \partial_t\bm A$ and the detector orientation ${\bm \zeta} (t)$ (green arrow).
    Also see Fig.~\ref{fig:signals}. Snapshot of the actual simulated field---leftmost panel---taken from~\cite{Amin:2022pzv}. For a movie of the vector field behaviour based on the simulations, visit \url{https://www.youtube.com/watch?v=bbw6yFRLS7s}.}
    \label{fig:Ellipse}
\end{figure}

\noindent{\it Equipartion and Vector Field Ellipses}: There exist various production mechanisms where disparate amounts of longitudinal (spin-$0$) and transverse (spin-$1$) helicities are produced~\cite{Graham:2015rva,Agrawal:2018vin,Dror:2018pdh,Long:2019lwl}. That is, for every ${\bm k}$, there could be different amounts of longitudinal and transverse components of $\bPsi_{\bm k}$. However, owing to non-linear gravitational clustering, such disparity within the two sectors is expected to have disappeared by now within our local cosmic vicinity. With non-linear gravitational clustering, we expect virialization, leading to the \textit{equipartition} of energy within all three degrees of freedom (dof). This would result in $2/3$ of the total power being contained within the (two) transverse dof and the remaining $1/3$ within the longitudinal ones. In this case, the vector field at each point (within a coherence region), which is formed out of a sum over a large number of Fourier modes, roughly traces out a {\it randomly oriented ellipse} (as opposed to oscillating along some fixed direction). This can be seen by noting that within time scales and length scales much smaller than the coherent ones, the spin current~\cite{Jain:2021pnk, Amin:2022pzv} is negligible since it scales with the typical speed ${\sigma}$. Hence, the local spin density (given by $\bm{s}={\bm A}\times\dot{\bm A}$) is conserved, implying that the local field vector ${\bm A}$ can execute a two-dimensional sinusoidal motion in general, i.e. it sweeps an ellipse. See~\cref{fig:Ellipse} for a visualization and a description of this evolution, and visit this \href{https://www.youtube.com/watch?v=bbw6yFRLS7s}{webpage} for a video. Such elliptical motion is appropriately used in, for example,~\cite{Fedderke:2021aqo, PPTA:2021uzb,Fedderke:2021rrm}.\\

{\it Random Linear Polarization}: Some studies make the assumption of linear polarization; i.e.~the vector field oscillates along a line in a fixed direction with this direction changing randomly every coherence time (see for example~\cite{Caputo:2021eaa,Frerick:2023xnf,Beaufort:2023qpd,Sun:2024qis}). However, as we argued above based on equipartition, an elliptical motion of the vector field is what one should expect. Nevertheless, we note that a preference for linear (circular) polarization could be generated by allowing for a non-gravitational attractive (repulsive) self-interaction \cite{Zhang:2021xxa,Jain:2022kwq, Jain:2022agt}. While these works argue for such a preference in isolated soliton-like configurations, whether a significant preference for linear (circular) polarization within each coherence patch can be achieved dynamically remains to be seen.\\

\noindent{\it Fixed Direction}: There could exist a misalignment production mechanism for vectors where the entire observable Universe (or at least a large portion of it) has the vector field oscillating in a fixed direction~\cite{Arias:2012az,Alonso-Alvarez:2019ixv}. Such setups indeed lead to a fixed direction of oscillation of the vector field (apart from randomly oriented small perturbations). This would be distinct from the equipartition case we consider. However, there are difficulties associated with their production mechanisms~\cite{Himmetoglu:2008zp,Himmetoglu:2009qi,Karciauskas:2010as}. 

\begin{figure}
    \centering
    \includegraphics[width=6in]{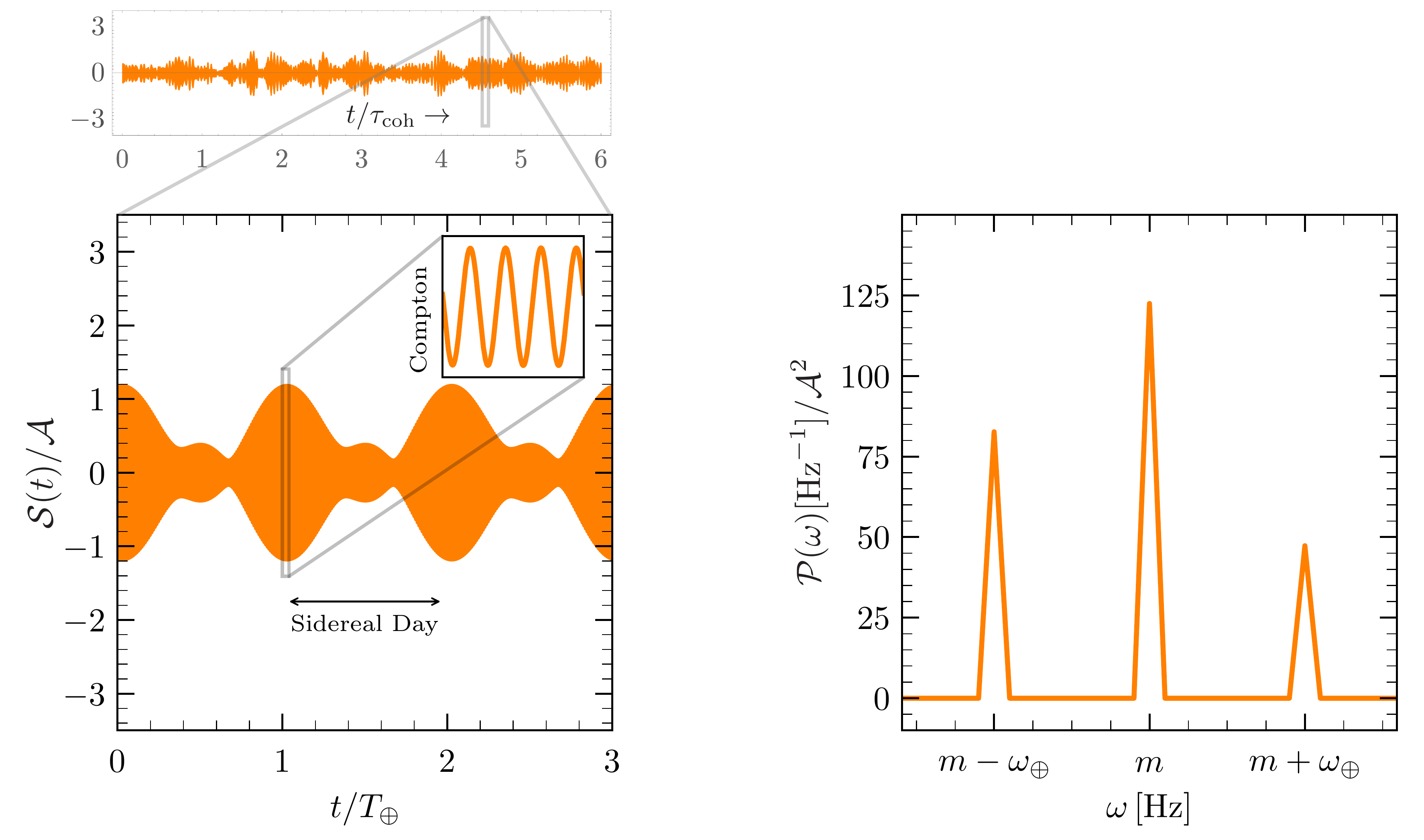}
    \caption{An example of the expected signal $\mathcal{S}(t)\approx g {\bm \zeta}(t)\cdot\partial_t\bm A(t)$. This depends on the shape, orientation and period of the vector field ${\bm A}(t)$, as well as the detector axis $\bm{\zeta}(t)$, which rotates with the Earth, and a model-dependent coupling $g$. Also refer to Fig.~\ref{fig:Ellipse}. We show the signal in the time (\textbf{left}) and Fourier (\textbf{right}) domains. For the time domain signal, we show its behaviour over several coherent times (\textbf{upper}) and sidereal days (\textbf{lower}), the latter of which is the expected observed signal in our sensor. The inset shows the oscillation of the signal over several Compton times. Note the appearance of three peaks: a single Compton peak at frequency $m$ in the middle and two additional ones appearing due to Earth's rotation---a \textit{difference} peak at $d \equiv m - \omega_\oplus$ and a \textit{sum} peak at $s \equiv m + \omega_\oplus$. The time domain signal is simulated from \cref{eq:low-t-sig}, the periodogram of which is taken via \cref{eq:periodogram-dft}. For the simulation, we have used the observation time  $T_\mathrm{obs} = 10 T_\oplus$, and the results are expressed in terms $\mathcal{A} \equiv g\sqrt{2\rho / 3}$ (where $\rho$ is the local DM density). For additional details on the simulation, see text. The peaks are contained within bins whose widths correspond to the resolution of the frequency space data, given by $\Delta \omega = 2 \pi / T_\mathrm{obs}$. Note that the general elliptical behavior of the DM field allows for different power in the sum and difference peaks. In contrast, the linearly polarized DM field would lead to equal power in these peaks.}
    \label{fig:signals}
\end{figure}

%~~~~~~~~~~~~~~~~~~~~~~~~~~~~~~~~
\subsection{The Detector Signal}
%~~~~~~~~~~~~~~~~~~~~~~~~~~~~~~~~

Let ${\bm \zeta}(t)$ be the ``antenna" direction---the time series signal is then given by $\hat{\mathcal{S}}(t) = g\,{\bm \zeta}(t)\cdot\hat{\bm E}(t) \approx g\,{\bm \zeta}(t)\cdot\partial_{t}\hat{\bm A}(t)$. Here, $g$ is some overall coefficient containing the coupling constant and any other possible model parameters, and we have used the fact that the produced dark electric field is approximately given by ${\bm E} \approx \partial_t{\bm A}$ in the non-relativistic limit of the vector DM field. In this work, we will take the antenna to always point towards the zenith; however, we will comment on this assumption in~\cref{sec:stats}. With $\phi$ denoting the latitude (where $\phi=0^\circ$ and $\phi = 90^\circ$ respectively correspond to the equator and poles), and $\omega_{\oplus}$ denoting the angular rotation frequency of the Earth, we have ${\bm \zeta}(t) = (\cos\phi\cos(\omega_{\oplus}t),\, \cos\phi\sin(\omega_{\oplus}t),\, \sin\phi)^\intercal$. Then, the signal $\hat{\mathcal S}(t)$ is
\begin{align}
    \label{eq:low-t-sig}
    \hat{\mathcal S}(t)|_{T \ll \tau_{\rm coh}} \approx g\,&\sqrt{2\rho / 3}\Bigl[\hat{\alpha}_x \cos\phi \cos\left(m t + \hat{\varphi}_x\right)\cos(\omega_{\oplus}t) + \hat{\alpha}_y\cos\phi\cos\left(m t + \hat{\varphi}_y\right)\sin(\omega_{\oplus}t) \nonumber \\    
    & \quad~ + \hat{\alpha}_z \sin\phi \cos\left(m t + \hat{\varphi}_z\right)\Bigr]\,.
\end{align}
Ultimately, we perform our analysis in Fourier space, considering the (one-sided) periodogram generated by \cref{eq:low-t-sig}. This is proportional to the mod-square of the discrete Fourier transform of $\hat{\mathcal{S}}(t)$ and is given by
\begin{equation}
    \label{eq:periodogram-dft}
    \hat{\mathcal P}(\omega) \equiv 2 \frac{(\Delta t)^2}{T_\mathrm{obs}} \Bigg| \sum_{n = 0}^{N - 1} \hat{\mathcal S}(t_n) e^{i \omega n \Delta t} \Bigg |^2\,.
\end{equation}
Here, $\omega$ is the angular frequency, $N$ is the number of points sampled in the time domain, and $\Delta t \equiv T_\mathrm{obs} / N$ is the sampling frequency. The factor of two accounts for the `folding' of the result from negative angular frequencies to positive angular frequencies, producing the one-sided periodogram that ignores the former. We define the dimensionless parameter
\begin{equation}
    \beta \equiv \sqrt{\frac{\mathcal{A}^2 T_\mathrm{obs}}{2 \sigma^2}}\,, \quad{\rm with}\quad \mathcal{A} \equiv g\sqrt{2\rho / 3}\,,
    \label{eq:beta}
\end{equation}
where $\sigma$ is the noise power spectral density (PSD). Typically, $\mathcal{A}$ is an acceleration or a force for accelerometer studies. With these definitions, the signal periodogram, normalised by the noise PSD (which we call the excess power $\hat{\lambda}$), is given by
\begin{equation}
\label{eq:low-t-period}
\begin{split}
\hat{\lambda}(\omega) \equiv \frac{\hat{\mathcal P}(\omega)}{\sigma^2} = \frac{\beta^2}{4}\Big\{&\left[\hat{\alpha}_x^2 + \hat{\alpha}_y^2 + 2 \hat{\alpha}_x \hat{\alpha}_y \sin(\hat{\varphi}_y - \hat{\varphi}_x)\right]\cos^2\phi\,\delta_{\omega,\,s}\\ 
+&\left[\hat{\alpha}_x^2 + \hat{\alpha}_y^2 - 2 \hat{\alpha}_x \hat{\alpha}_y \sin(\hat{\varphi}_y -\hat{\varphi}_x)\right]\cos^2\phi\,\delta_{\omega,\,d}\\
+&~4 \,\hat{\alpha}_z^2 \sin^2\phi\,\delta_{\omega,\,m}\Big\}\,.
\end{split}
\end{equation}
Here, we have defined the `sum' and `difference' angular frequencies, $s \equiv m + \omega_\oplus$ and $d \equiv m - \omega_\oplus$, respectively. The $\delta_{a,\,b}$ are Kronecker delta functions over angular frequencies.

We note that $\sigma$ can be frequency dependent. This means that the expected noise PSD within each signal-containing bin can be different. However, for the remainder of this work, we take $\sigma$ to be approximately constant, which is a good approximation in the frequency range of the signal, $\Delta \omega \sim \omega_\oplus \equiv 2\pi / T_\oplus \approx \SI{7.3e-5}{\hertz}$, where $T_\oplus = \SI{23.93}{\hour}$ is the Earth's sidereal period. We comment on this approximation further in \cref{sec:application} when we consider a concrete sensor and DM model.

\cref{fig:signals} shows an example of a signal in the short observation time regime, in both the time (\cref{eq:low-t-sig}) and Fourier (\cref{eq:low-t-period}) domain. To generate them, we have taken  $\mathcal{A} = 1\,[\mathcal{A}]$, ${m = 2\pi\,\si{\hertz}}$, $\phi = 45^\circ$, $T_\mathrm{obs} = 10 T_\oplus$, and, for the purposes of fast convergence, $T_\oplus = \SI{100}{\second}$.  Here, $[\mathcal{A}]$ are the units of $\mathcal{A}$, which depend on the quantity being measured by the experiment. We have also taken $\bvec{\alpha} \equiv (\alpha_x,\,\alpha_y,\,\alpha_z)^\intercal  = (1,\,0.7,\,0.2)^\intercal$ and $\bvec{\varphi} \equiv (\varphi_x,\, \varphi_y,\, \varphi_z)^\intercal = (\pi/2,\, \pi/4,\, \pi/3)^\intercal$. When running our future simulations, we sample these six variables independently from their respective distributions.

There are three characteristic timescales within the signal: the Compton scale, the Earth's rotation period, and the coherent (de Broglie) scale. The first two of these are present in the larger panels of ~\cref{fig:signals}. The Earth's rotation period is evident from the time domain signal, which we have shown for three full rotation periods. The Compton scale is much faster than this scale (see inset), making the signal appear solid in shape. Crucially, we see that the vector ULDM field leaves a characteristic three-peak signal in the Fourier domain\footnote{Such a signal would also be seen in the case of the gradient of a scalar field, as mentioned in Ref.~\cite{Lisanti:2021vij}, and its form is similar to that of \cref{eq:low-t-sig}. See~\cref{app:likelihood_gradientscalar} for details.}. One peak is present at the Compton frequency $\omega = m$, which previous frequency-space analyses have focused on \cite{Carney:2019cio, Manley:2020mjq}. However, a further two peaks manifest as a result of the Earth's rotation, which are spaced $\omega_\oplus$ away from the Compton peak. These additional peaks, the use of which has been ignored in previous accelerometer analyses\footnote{Ref.~\cite{Carney:2019cio} also noted this spectral splitting; however, the focus was on the Compton peak for the purposes of that analysis. In Ref.~\cite{Fedderke:2021rrm}, this splitting as it relates to a global network of magnetometers was used.}, only appear as $T_\mathrm{obs} \geq T_\oplus$; shorter observation times do not give us enough resolution in the frequency domain to resolve them\footnote{For us, the observation time is defined as the amount of time that data are continuously taken. As such, the total expedition time of the entire experiment is equivalent to $T_\mathrm{obs}$.}. We call the peak at the Compton frequency the \textit{Compton} peak, that at $s$ the \textit{sum} peak, and that at $d$ the \textit{difference} peak.

We argued earlier that, within a coherence patch, we expect the vector field to undergo an elliptical motion with period $2\pi/m$ (see \cref{fig:Ellipse}), as opposed to a linear one commonly used in the literature. In both the linear and elliptical cases, the time domain signal, $\mathcal{S}(t)$, is sinusoidal and contains the angular frequencies $m$ and $m \pm \omega_\oplus$. Repeating the analysis of~\cite{Caputo:2021eaa} in the time domain, but without the linear polarization assumption, we expect qualitatively similar results (with more statistical spread on the time averaged power). However, there are some important differences when analysing the expected signal in Fourier space. 

In the elliptical case, the power contained in the $m$ and $m\pm\omega_{\oplus}$ peaks is statistically uncorrelated (see~\cref{app:deriv-marginal}). On the other hand, for the linear polarization case, the power at $m$ and $m\pm \omega_{\oplus}$ is correlated, with equal power in the sum and difference peaks. This can be seen by noting that, in this case, all the components of the vector are in phase. The statistical independence in the elliptical case significantly simplifies our analysis pipeline for projected sensitivities. Furthermore, the distinction in power at $m\pm \omega_\oplus$ is also relevant in case of a detection since we would expect different powers in the elliptical case. We show the statistical independence arising from the elliptical case in \cref{app:deriv-marginal} and the statistical dependence of the sum and difference peaks arising from the linear polarization case in \cref{app:likelihood_linearpol}.

\section{Statistical Analysis}
\label{sec:stats}

We now consider the projected exclusion limits that a generic experiment would be able to set using our three-peak analysis. To do this, we use a series of likelihood-ratio tests. 

\subsection{Signal Likelihood}
\label{subsec:sig-lik}

For our likelihood, we follow a hybrid frequentist-Bayesian approach, defining a marginalized likelihood in which all nuisance parameters are integrated out. In our case, these are the random Rayleigh parameters, $\bvec{\alpha}$, and random uniform DM phases, $\bvec{\varphi}$. Such a hybrid approach has already been used in the context of ultralight bosonic dark matter \cite{Centers:2019dyn,Nakatsuka:2022gaf}. Our work differs from Ref.~\cite{Centers:2019dyn} since they focused on an axion-like signal as opposed to that from vector DM. It goes beyond Ref.~\cite{Nakatsuka:2022gaf} since they did not consider the peaks arising from the rotation of the Earth in their analysis.%\MA{check again}

The full likelihood in Fourier space is well-known to follow a non-central $\chi^2$ with two degrees of freedom \cite{1975ApJS...29..285G}. In our case, the non-centrality parameter is the total signal amplitude in \cref{eq:low-t-period}. The marginalized likelihood is then given by
\begin{equation}
    \label{eq:lik-full}
    \mathcal{L}_\mathrm{marg}(\beta,\,\phi;\,p) = \int \mathrm{d}^3 \bvec{\alpha}\, \mathrm{d}^3 \bvec{\varphi}  \left[\chi_\mathrm{nc}^2\left(p;\,\, k=2,\, \lambda(\beta,\,\phi,\,\bvec{\alpha},\,\bvec{\varphi})\right)\,\Pi (\bvec{\alpha},\,\bvec{\varphi})\right]\,,
\end{equation}
where $\Pi$ describes the priors of our random parameters and $p$ is the random variable we expect to measure in an experiment. We can express the result of \cref{eq:lik-full} completely analytically and provide a full derivation of it in \cref{app:deriv-marginal}, only quoting the final result here. The likelihoods in the signal-containing bins, which we call the Compton and sum/difference likelihoods, are respectively
\begin{equation}
\label{eq:liks}
\begin{split}
\mathcal{L}_m(\beta,\,\phi;\,p) &= \frac{1}{2 + \beta^2\sin^2\phi} \exp\left[-\frac{p}{2 + \beta^2\sin^2\phi}\right]\,, \\
\mathcal{L}_{s/d}(\beta,\,\phi;\,p) &= \frac{1}{2 + (\beta^2\cos^2\phi)/2} \exp\left[-\frac{p}{2 + (\beta^2\cos^2\phi)/2}\right]\,.
\end{split}
\end{equation}
Note that, when $\beta = 0$, we correctly retrieve central $\chi^2$ distributions in each bin, corresponding to the background-only case. The result for the Compton peak matches that of Refs.~\cite{Centers:2019dyn,Nakatsuka:2022gaf}. The result for the sum/difference peaks is new. For completeness, we also derive the equivalent of \cref{eq:liks} for the linear polarization case in \cref{app:likelihood_linearpol}.

\begin{figure}
    \centering
    \includegraphics{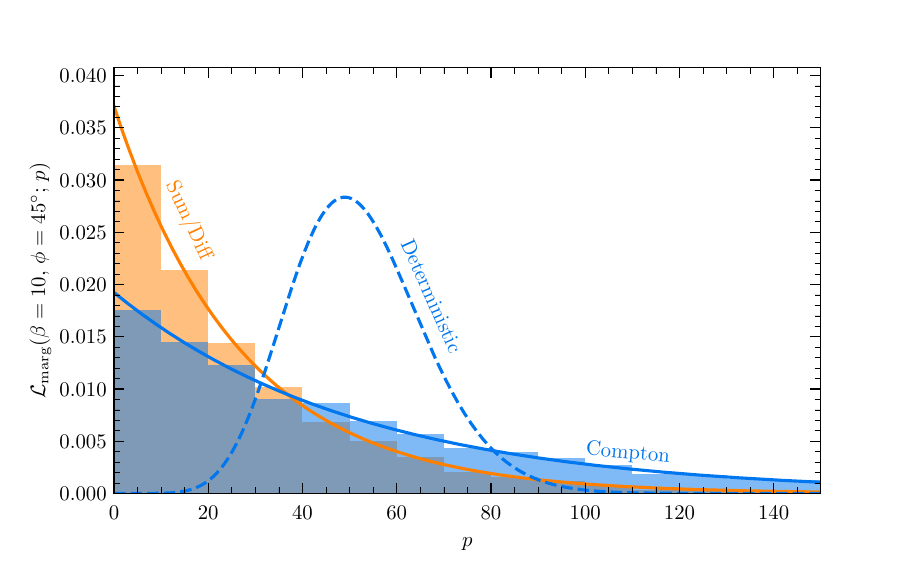}
    \caption{Example likelihoods for each of the three signal peaks once the stochastic variables have been marginalised over. The bars show the result of a numerical simulation of the noise-normalised periodogram values beginning from \cref{eq:low-t-sig}, while the solid lines show the analytical result of \cref{eq:liks}. Here, $\phi$ is the latitude of the experiment, $p$ (a random variable) is the value of the measured excess power, and $\beta$ is defined as per \cref{eq:beta}. Also shown as a dashed line is the deterministic result for the Compton peak, where the stochastic variable $\alpha_z^2$ is set to its expectation value, $\langle\alpha_z^2\rangle = 1$.}
    \label{fig:marginal-lik}
\end{figure}

In \cref{fig:marginal-lik}, we show a comparison between a numerical simulation of the signal likelihoods and the analytical results of \cref{eq:liks}. For the former, we begin from \cref{eq:low-t-sig}, simulating $10^6$ realisations of the time-domain signal. We take $\mathcal{A} = 1\,[\mathcal{A}]$, $T_\mathrm{obs} = 10 T_\oplus$, $\sigma^2 = 5\,[\mathcal{A}]^2\,\mathrm{Hz^{-1}}$, and $\phi = 45^\circ$. For the purposes of fast convergence in our simulations, we take $m = 2\pi\,\si{\hertz}$ and $T_\oplus = \SI{100}{\second}$. For each run, we sample $\bvec{\alpha}$ and $\bvec{\varphi}$ from independent Rayleigh and uniform distributions respectively, as given by \cref{eq:distributions}. To generate the time domain signal, we add Gaussian-distributed white noise with zero mean and variance given by $\sigma^2_t = \sigma^2 \Delta t$. Finally, we compute the periodogram of the signal following \cref{eq:periodogram-dft}, dividing by $\sigma^2$ to produce the excess power in each frequency bin. The resulting normalised distribution of $p(\omega)$ for the Compton bin ($\omega = m$) and the sum/difference bins ($\omega = m \pm \omega_\oplus$) are in excellent agreement with our analytical result. We also show the likelihood governing the deterministic result for the Compton peak: a non-central $\chi^2$ with two degrees of freedom and non-centrality parameter given by the last term of \cref{eq:low-t-period}. Without the nuisance parameter $\alpha_z^2$ integrated out, we instead set it to its expectation value: $\langle\alpha_z^2\rangle = 1$. We see that higher values of $p$ are favoured in this case, which would ultimately lead to an overly aggressive constraint on $\beta$.

The full likelihood over all frequency space is then given by the product of the likelihoods in each frequency bin,
\begin{equation}
    \mathcal{L}(\beta,\,\phi;\,\bvec{p}) = \prod_{i=1}^{N_\mathrm{bins}} \mathcal{L}_\mathrm{marg}(\beta,\,\phi;\,p_i)\,,
    \label{eq:lik-tot}
\end{equation}
where $p_i$ represents the excess power density in the $i^\mathrm{th}$ frequency bin, $\bvec{p}$ is the full data vector, and the product runs over all $N_\mathrm{bins}$ frequency bins. Ultimately, since our signal only manifests in three bins, it suffices for us to consider only those bins that could potentially contain a signal, and we may ignore all other bins. 
We can express the likelihood in this way because each bin is statistically uncorrelated, as we show in \cref{app:deriv-marginal}. This is in contrast with the analysis performed in Ref.~\cite{Lisanti:2021vij}, where a similar study was conducted in the case of the gradient of a scalar in the time domain. There, a complicated covariance matrix had to be computed to account for correlations in the signal at different times. In Fourier space, these covariances disappear. The power of performing this analysis in the frequency domain is thus not only that the signal is contained within a small number of bins, but also that these bins are statistically independent, which allows us to treat the statistics in a significantly simpler way. 

Crucially, once the latitude of the experiment, $\phi$, is fixed, the likelihood depends on the product of all experimental variables via the dimensionless parameter $\beta$. This means that we can set a more holistic limit that is independent of the specifics of an experiment. Once the form of $\mathcal{A}$ (which depends on both the experiment and the DM model), the observation time $T_\mathrm{obs}$, and the noise profile $\sigma$ are known, the ensuing limit on $\beta$ can be recast to one on the model parameters of interest. This makes our analysis, both the results and overall logic, as generally useful as possible. 

\subsection{Projected Exclusions}
\label{subsec:proj-excl}

To derive our limits, we construct the one-sided log-likelihood-ratio test statistic, defined as
\begin{equation}
\label{eq:test-stat}
q_\mathcal{\beta} \equiv 
\begin{cases}
-2 \ln\left[\frac{\mathcal{L}(\beta,\,\phi;\,\bvec{p})}{\mathcal{L}(\hat{\beta},\,\phi;\,\bvec{p})}\right] \quad &\text{if} \quad \hat{\beta} \geq \beta \\
0 \quad &\text{else}
\end{cases}
\,,
\end{equation} 
where $\hat{\beta}$ is that value of $\beta$ which maximises the likelihood given the observed data set $\bvec{p}$, characterising the best-fit model. This statistic is defined as a piecewise function as we only expect excess signals to be disfavoured when excluding a value of $\beta$ in a one-sided test. This corresponds to values of $\beta$ greater than the best-fit value. Values below this are deemed under-fluctuations and considered consistent with observation. This statistic tells us how consistent the data are with a signal defined by $\beta$ compared to the best-fit model, with  zero representing perfect consistency and large values indicating high inconsistency.

The general idea is that we want to exclude those values of $\beta$ that lead to excessive values of $q_\beta$. We do this by considering the distribution $f(q_\beta)$ that we expect to arise when many hypothetical experiments perform a measurement. Generating the $\alpha_\mathrm{conf} \%$ confidence level (CL) limit then depends on finding the value of the test statistic, $q_\beta^\mathrm{lim}$, for which an experiment has an $\alpha_\mathrm{conf} \%$ probability of attaining that value or below. That is, we solve for $q_\beta^\mathrm{lim}$ in
\begin{equation}
    \label{eq:int-qlim}
    \int_0^{q_\beta^\mathrm{lim}} f(q_\beta)\,\dl q_\beta = \alpha_\mathrm{conf}\,.
\end{equation}
From $q_\beta^\mathrm{lim}$, we then find that $\beta^\mathrm{lim}$ which gives us this value for the test statistic. This is our $\alpha_{\rm conf} \%$ CL limit on our parameter of interest, $\beta$. 

Ultimately, we compute $90
\%$ CL limits ($\alpha_{\rm conf} = 0.90$) via a series of Monte Carlo (MC) simulations, following the above procedure in each MC run\footnote{While asymptotic formulae exist for computing exclusion limits, we have opted for an MC approach to ensure proper coverage \cite{Cowan:2010js}. Since we are following a hybrid frequentist-Bayesian approach, we found this to be an especially pertinent check.}. For a given choice of $\beta$, we begin by simulating the distribution $f(q_\beta)$. We do this by simulating $10^6$ experiments, sampling the data $p$ for each signal bin directly from the verified likelihoods given in \cref{eq:liks}. In each run, we find $\hat{\beta}$ and compute the distribution of $q_\beta$ for a range of $\beta$ and $\phi$ values. We then fit our distributions to the ansatz
\begin{equation}
\label{eq:f-ansatz}
f(q_\beta) = (1 - \varpi)\,\delta(q_\beta) + \varpi\,\chi^2(q_\beta;\,k = 1),\,
\end{equation}
where $\varpi \in [0,\,1]$ is a scale factor ensuring that the distribution is normalised to $1$. This ansatz is inspired by the asymptotic result of Chernoff (where $\varpi = 1/2$), which itself is a limiting case of Wilk's theorem when the true value of $\beta$ lies on the boundary of its domain (which is true in our case, since our background-only data set if defined by $\beta = 0$) \cite{Chernoff:1954eli,Algeri:2019lah}. For each run, we find the best-fit value of $\varpi$. From \cref{eq:f-ansatz}, we can then invert \cref{eq:int-qlim} to find $q_\beta^\mathrm{lim}$. For the $\alpha_\mathrm{conf} \%$ CL limit, we have that
\begin{equation}
\label{eq:q_beta_lim}
q_\beta^\mathrm{lim} = 2\left[\mathrm{Erf}^{-1}\left(1 + \frac{\alpha_\mathrm{conf} - 1}{\varpi}\right)\right]^2\,.
\end{equation}
In generating our distributions, we find a weak dependence on the chosen values of $\beta$ and $\phi$. However, this leads to a small change in the ultimate value of $q_\beta^\mathrm{lim}$. We take the mean value of it over our fits as its best estimator, yielding $q_\beta^\mathrm{lim}\simeq2.43$, and use this throughout the rest of our study. Note that this is close to the asymptotically expected result of $q_\beta^\mathrm{lim}\simeq2.71$ for a 1 d.o.f. problem \cite{Cowan:2010js}. We could have derived a similar result by integrating the resulting $q_\beta$ histograms; however, we attain a good fit to \cref{eq:f-ansatz}, and it provides us with a closed-form solution for $q_\beta^\mathrm{lim}$, as per \cref{eq:q_beta_lim}.

\begin{figure}[t!]  
\centering
\includegraphics{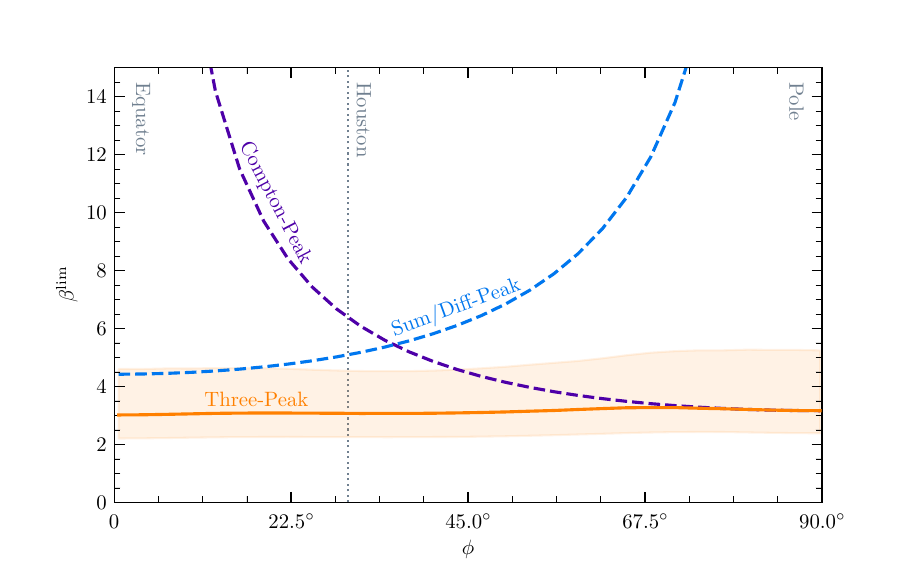}
\caption{The $90\%$ CL limits ($\beta^\mathrm{lim}$) on the dimensionless parameter $\beta \equiv \sqrt{\mathcal{A}^2 T_\mathrm{obs} / (2 \sigma^2)}$ (see \cref{eq:beta}) derived from our MC analysis . We show the results of our three-peak analysis and those of two single-peak analyses focusing on the Compton peak and on one of the sum or difference peaks. The shaded region indicates the $1\sigma$ error bar on our three-peak analysis. The dotted line indicates the latitude of Houston, which we use in \cref{sec:application}. Our three-peak analysis generally provides a better/comparable limit to either of the two single-peak analyses and is largely latitude-independent.}
\label{fig:beta-lim}
\end{figure}    

With $q_\beta^\mathrm{lim}$ at hand, we may derive the main result of this section, $\beta^\mathrm{lim}$. We once again follow an MC approach, generating $10^6$ data sets consistent with a background-only observation (setting $\beta = 0$), and finding, for each hypothetical experiment, that value of $\beta$ for which \cref{eq:test-stat} returns $q_\beta = q_\beta^\mathrm{lim}$. This produces a distribution of limits, for which we take the median as our best estimator. We also produce the $1\sigma$ error bars on our limits by finding the $16^\mathrm{th}$- and $84^\mathrm{th}$-percentile of the distribution in $\beta^\mathrm{lim}$.

We show the $90\%$ CL limit arising from our three-peak analysis in \cref{fig:beta-lim}. Also shown are the corresponding results from two single-peak analyses focusing solely on the Compton and either one of the sum or difference peaks. These results follow the same MC procedure as above but take as the full likelihood only the Compton or the sum/difference likelihood given in \cref{eq:liks}. 
We see that the three-peak analysis produces a limit that is largely latitude-independent, rising slightly towards the pole. This is because the sensitivity axis at this latitude only has a component parallel to the Earth's rotation axis and is thus only able to pick out the Compton peak. 

This latitude-independent limit is in contrast with the analyses that focus on only single peaks, which are both highly sensitive to where the experiment is placed. For the study focusing on the Compton peak, the constraining power is optimal at the pole, where all of the power is contained in the Compton frequency bin, and it rapidly declines towards the equator, where the Compton peak disappears. Note that this and the three-peak results join at this point, with no difference between the approaches. Conversely, for an analysis focusing on one of the sum or difference peaks, the situation is the opposite. In this case, the results of this and the three-peak methods do not converge since half the power is contained in a single one of the sum or difference peaks at the equator; the three-peak analysis captures all of this power, whereas the single-peak analysis misses half of the power. Thus, the strength of our analysis is that the constraining power is retained no matter where an experiment is placed, such that its latitude is rendered largely irrelevant from the viewpoint of constraining a ULDM signal. 

We emphasize that the interpretation of~\cref{fig:beta-lim} is as a set of exclusion lines whereby background pseudodata is generated and the assumed DM signal strength is constrained. The key point is that the level of this constraint depends on the assumption one makes on the nature of the DM signal given a non-detection. Taking this signal to be only a single peak in Fourier space then leads to constraints that are generally dependent on the latitude of the experiment, rapidly weakening towards latitude extremes. On the other hand, employing the full signal model consisting of three peaks yields stronger constraints that are almost independent of the experiment placement.

Throughout the above analysis, we assumed that the sensitivity axis pointed in the zenith direction. However, we can relax this assumption and consider what our results would look like if this axis pointed in some different direction, say for instance directions perpendicular to the zenith---namely, the East/West and North/South directions\footnote{Here we assume that the sensitivity axis of the detector is locally fixed with respect to the zenith.}. If North/South-pointing, all of the curves in \cref{fig:beta-lim} would be flipped about the line $\phi = 45^\circ$. While the strongest limit for the Compton peak would now occur at the equator instead of the pole (and vice versa for both the sum and difference peaks), crucially we would still retain a largely latitude-independent constraint for our three-peak analysis. This is because, throughout the experiment, the directionality of the detector makes a cone, making it sensitive to the vector DM power in all the three directions. If, instead, the axis pointed towards the East/West, we would only see the sum and the difference peaks. This is because, throughout the experimental expedition, the directionality of the detector is restricted to lie on a plane (which would necessarily be perpendicular to the rotation axis of the Earth). Therefore, it is always insensitive to the power contained across the perpendicular direction, which is tied to the standalone Compton peak. One final possibility is when the directionality of the detector traces out a line throughout the experiment. This is only possible when it points parallel to the Earth's rotation axis (at any given latitude). In this case, we would naturally be oblivious to the Earth's rotation and hence to the sum and difference peaks, and we would only be able to resolve the Compton peak.

In summary, the best-case scenario is when the detector's sensitivity axis traces out a cone. In this case, we capture all the three peaks since we
are sensitive to the vector DM power across all three directions (Earth's rotation axis and the two orthogonal directions). 

For comparison, we also derive the scaling relationship of the limits on $\beta$ under the assumption of linear polarization in \cref{app:likelihood_linearpol} using 
a simpler two-peak Asimov analysis. We find that the limits are mostly affected towards the equator, with the largest scaling factor being $\sim 1.2$. This difference becomes larger for higher desired confidence levels.

\section{Application to Accelerometer Studies}
\label{sec:application}

As an application of our analysis strategy, we consider a concrete sensor and DM model. As our sensor, we take the canonical optomechanical light cavity, which can be used to perform acceleration measurements by continuously measuring the distance between fixed and movable cavity mirrors. As our model, we consider `dark photon' DM stemming from a gauged $U(1)_{B - L}$ symmetry, leading to wavelike DM in the ultralight regime that couples to the difference between baryon number, $B$, and lepton number, $L$. Gauging such a charge is popular in the context of particle physics, since it naturally leads to the introduction of right-handed neutrinos and, hence, can account for the non-zero neutrino masses \cite{Peskin:1995ev,Basso:2008iv,Kanemura:2014rpa}. Motivation for such an ultralight gauge boson can also be found in \cite{Fayet:1980ad,Fayet:1980rr}. The associated Lagrangian density reads
\begin{equation}
\label{eq:lagran}
    \mathcal{L} \supset -\frac{1}{4} A'^{\mu \nu}A'_{\mu \nu} - \frac{1}{2} \m^2 A'^{\mu}A'_{\mu} - g_{B - L} j_{B - L}^\mu A'_\mu\,,
\end{equation}
where $A'^{\mu\nu} \equiv \partial^\mu A'^\nu - \partial^\nu A'^\mu$ is the field strength tensor for the new field, $\m$ is the mass of the field, and $j^\mu_{B-L}$ is the $B-L$ vector current. Explicitly, it is given by
\begin{equation}
    j^\mu_{B-L} = \sum_f Q_{B - L}^f \bar{f} \gamma^\mu f\,,
\end{equation}
where the sum runs over all fermions in the SM and where $Q_{B - L}^f$ is the $B - L$ charge of the fermion $f$.
A multitude of studies have considered this combination and set limits or projections on $B - L$ coupled DM \cite{Carney:2019cio,Manley:2020mjq,Nakatsuka:2022gaf,LIGOScientific:2021ffg,Abe_2021,Badurina:2019hst}. 

However, those works that performed a Fourier space
analysis only had access to the Compton peak. This is because the total experiment integration times had to be such that $T_\mathrm{obs} \sim \SI{1}{\hour} < T_\oplus$ to maintain experimental stability, dictated by retaining the coherence of the laser in optomechanical cavity setups. In the event that one can instead measure for at least $\sim \SI{1}{day}$, other peaks can be resolved. As we discussed in \cref{sec:stats}, an analysis for an axial sensor that then does not account for these additional peaks is sub-optimal, as it fails to capture the full signal and therefore suffers from a signal loss at a range of latitudes. Moreover, ignoring the randomness of the nuisance variables leads to an overly aggressive constraint, as was illustrated in \cref{fig:marginal-lik}. Our more holistic three-peak strategy, which also considers this stochasticity, retains the full signal and is largely latitude-independent in its constraining power. Therefore, this choice of sensor and DM model makes for an excellent case study with which to showcase the improved constraining power of our method.

Furthermore, a proper vector treatment of the DM field in Fourier space, which includes the stochasticity of the ULDM field variables and the effect of the rotation of the Earth, has not been done. Ignoring the randomness of the nuisance variables leads to an overly aggressive constraint, as was also pointed out in Refs.~\cite{Centers:2019dyn,Nakatsuka:2022gaf}.

\subsection{Recasting Generalised Limits onto $B - L$ Dark Matter}

To recast our limits on $\beta$ shown in \cref{fig:beta-lim} into one on the parameter of interest for this model, the gauge coupling strength $g_{B - L}$, we must define four quantities. Firstly, we must make clear what the quantity $\mathcal{A}$ is for this experiment, which will depend on both the model and the signal of interest for this sensor. Secondly, we must choose an observation time, $T_\mathrm{obs}$. Thirdly, we must elaborate on what the concrete noise profile, $\sigma(\omega)$, for this type of experiment is. Lastly, and least importantly following from our discussion above, we must choose a latitude for the experiment. Once these are known, \cref{eq:beta} can be re-arranged for $g_{B - L}$, giving us our model- and experiment-specific $90\%$ CL limit.

For $B - L$ coupled dark photon dark matter, the relevant signal is a differential acceleration. This is given by \cref{eq:low-t-sig} in the time domain, with a now concrete choice for $\mathcal{A}$,
\begin{equation}
    \mathcal{A} \equiv g_{B - L} \Delta_{ij} a_0\,.
    \label{eq:amp-bl}
\end{equation}
Here, $g_{B - L}$ is the gauge coupling strength of the model, $\Delta_{ij}$ is the differential $B -L$ charge per nucleon between materials $i$ and $j$, given by
\begin{equation}
    \Delta_{ij} \equiv \bigg| \frac{Z_i}{A_i} - \frac{Z_j}{A_j} \bigg|\,,
\end{equation}
and $a_0 \equiv \sqrt{2 \rho / 3} \, u^{-1} \simeq \SI{e12}{\meter\per\second\squared}$ is a characteristic acceleration imparted by the field to each nucleon (with $u$ being the atomic mass unit). For most materials, $\Delta_{ij} \sim 0.1$, which we take in the following analysis \cite{Carney:2019cio}. Note that, for this particular model, $g \equiv g_{B - L} \Delta_{ij} / u$.

For our observation time, we take $T_\mathrm{obs} = 10 T_\oplus$ to be firmly in the regime where the three peaks can be resolved. For light cavities, which may only be able to remain coherent over the scale of hours rather than days, such a runtime is optimistic. However, since our aim here is merely to showcase how our method can be used concretely and compare to the works of Refs.~\cite{Carney:2019cio,Manley:2020mjq}, we do not see this as an issue. Our strategy is general and can be applied to any axial sensor and vector-like DM model, and we have settled on this configuration only for the sake of argument; other sensor technologies, such a magnetically levitated sensors, do not have this issue. Moreover, multiple cavities or data-stacking techniques can be employed to mitigate this issue. 

We model the background according to Refs.~\cite{Carney:2019cio,Manley:2020mjq}. Namely, we split the total expected background PSD $\sigma^2$, which we will write as $S_{aa}$ as per convention, into a thermal, shot-noise, and back-action component,
\begin{equation}
	S_{aa} = S_{aa}^\mathrm{Th} + S_{aa}^\mathrm{SN} + S_{aa}^\mathrm{BA}\,.
\end{equation}
In what follows, we do not discuss the forms of these noise terms; we instead refer the reader to Ref.~\cite{Clerk:2008tlb,Beckey:2023shi} for a review on the topic. The thermal component is given by
\begin{equation}
	S_{aa}^\mathrm{Th} \equiv \frac{4 k_B T \gamma}{m_s}\,,
\end{equation}
where $\gamma$ represents the couplings between the sensor and the thermal bath of temperature $T$, $m_s$ is the mass of the sensor, and $k_B$ is Boltzman's constant. Typically, we parametrise the thermal coupling as $\gamma \equiv \omega_0 / Q$, where $\omega_0$ is the resonance frequency of the cavity and $Q$ is its quality factor. 
The measurement-added noise terms---the shot-noise and back-action noise---are respectively given by
\begin{equation}
	S_{aa}^\mathrm{SN}(\omega) \equiv \frac{\hbar \kappa L^2}{2\omega_L P_L} |\chi_c(\omega)|^{-2}|\chi_m(\omega)|^{-2}\,,
\end{equation}
and
\begin{equation}
	S_{aa}^\mathrm{BA}(\omega) \equiv \frac{2\hbar \omega_L P_L}{m_s^2 L^2 \kappa} |\chi_c(\omega)|^{2}\,.
\end{equation}
Here, $\kappa$ is the cavity loss, which quantifies the efficiency of the optical modes of the cavity, $L$ is the cavity length, $\omega_L$ is the angular frequency of the laser, and $P_L$ is its power. The mechanical susceptibility is given by
\begin{equation}
	\chi_m(\omega) \equiv \frac{1}{\omega_0^2 - \omega^2 + i \gamma \omega}
\end{equation}
and the cavity susceptibility by
\begin{equation}
	\chi_c(\omega) \equiv \frac{\sqrt{\kappa}}{i\omega - \kappa / 2}.
\end{equation}
Our choices for all of the above parameters except for the laser power, which we expand on below, are summarised in \cref{tab:config}. These are in keeping with the choices made in Refs.~\cite{Carney:2019cio,Manley:2020mjq}.

\begin{table}[t!]
	\centering
	\begin{tabular}{ll}
	\textbf{Quantity}	&  \textbf{Value} \\ \toprule[1pt]
	Sensor Mass ($m_s$)	&   \SI{10}{\milli\gram}\\
	Bath Temperature ($T$)	&  \SI{10}{\milli\kelvin} \\
	Quality Factor ($Q$)	& \num{e9} \\
	Cavity Loss ($\kappa$)	&   \SI{e6}{Hz}\\
	Laser Wavelength ($\lambda_L$)	& \SI{1.55}{\micro\meter} \\
	Cavity Length ($L$)	&  \SI{10}{\centi\meter} \\ 
    \bottomrule[1pt]
	\end{tabular}
	\caption{The optomechanical cavity configuration we have assumed in this work.}
	\label{tab:config}
\end{table}

The choice of where to tune the laser power is critical to give us competitive limits for a wide range of dark matter masses. We have found that we can achieve excellent limits for low dark matter masses, which are of most relevance to our work, by tuning the laser power such that the back-action and shot-noise components are minimised at low frequencies. That is, by finding that $P_L$ for which $\partial_{P_L}[S_{aa}^\mathrm{SN}(\omega \rightarrow 0) + S_{aa}^\mathrm{BA}(\omega \rightarrow 0)] = 0$ for a given choice of $\omega_0$. The required laser power to achieve this is given by
\begin{equation}
	P_L^\mathrm{min} = \frac{m_s \kappa^2 L^2 \omega_0^2}{8 \omega_L}\,.
\end{equation}
For the resonant frequencies we have considered, the laser power ranges from $P_L \sim \SI{e-12}{\W}$ for $f_0 = \SI{0.1}{\hertz}$ to $P_L \sim \SI{e-8}{\W}$ for $f_0 = \SI{10}{\hertz}$. 

We note that this choice of power tuning is different from the strategies usually employed in other studies. Typically, the laser power is tuned so that the measurement-added noise is either minimised on resonance or, as LIGO implements it, well above resonance \cite{LIGOScientific:2007fwp}. However, we found that both of these choices detriments the limit that we can draw at low masses, increasing the background at low frequencies beyond the thermal noise floor. For these other strategies to be beneficial at both the respective frequency targets and at low frequencies, the thermal noise would have to be significantly lower than that achieved with our choice of sensor mass, quality factor, and bath temperature.

In the treatment of our backgrounds, we have neglected seismic noise, which can become important at frequencies below $\sim \SI{10}{\hertz}$. Since we require the use of two materials to measure the differential acceleration peculiar to $B - L$ DM, we can envision constructing two sensors: one with a moveable mirror made of material $i$ and another of material $j$. By subtracting their signals, we both isolate the differential acceleration signature and remove backgrounds common to both---this includes the seismic noise component \cite{Graham:2015ifn}

Finally, we choose Houston as the location of our experiment, $\phi = 29.76^\circ$. From \cref{fig:beta-lim}, we find that $\beta^\mathrm{lim} \simeq 3.1$ for this latitude. However, we note that this information is almost unnecessary for the three-peak analysis since it is largely location-independent. We may then re-arrange \cref{eq:beta} accounting for \cref{eq:amp-bl} to give us our limit on ultralight $B-L$-coupled DM,
\begin{equation}
    g_{B - L} = \frac{\beta^{\rm lim}}{a_0 \Delta_{ij}}\sqrt{\frac{2S_{aa}}{T}}
    \label{eq:bl-coupling}
\end{equation}

\begin{figure}[t!]
    \centering
    \includegraphics{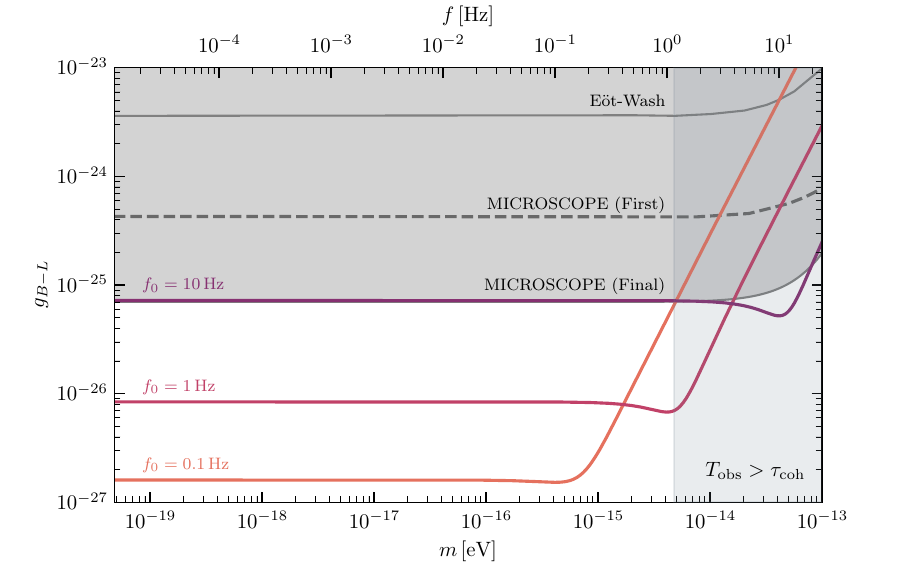}
    \caption{The $90\%$ CL limits on the gauge coupling for ultralight ${B - L}$ DM placed by an optomechanical cavity setup using our statistical framework. The limits using our three-peak analysis strategy (solid) for three resonance frequencies, $f_0 = \SIlist{0.1;1;10}{\hertz}$ are shown. Existing bounds from the E\"ot-Wash~\cite{Wagner:2012ui,AxionLimits} and MICROSCOPE experiments are shown in grey. For MICROSCOPE, we show the bound based on the first~\cite{Touboul:2017grn,Berge:2017ovy,MICROSCOPE:2019jix,AxionLimits} and final~\cite{MICROSCOPE:2022doy} results, the latter of which we compute in \cref{app:microscope}. The vertical shaded region indicates where the observation time $T_\mathrm{obs} = 10 T_\oplus$ becomes greater than the coherence time, where our framework is no longer valid. The top axis shows the Compton frequency for a given DM mass in \si{\hertz}.}
    \label{fig:b-l}
\end{figure}

We show our limits in \cref{fig:b-l} for three choices of resonance frequency: \SIlist{0.1;1;10}{\hertz}. For all but the last of these frequencies, we are able to exclude new regions of the UDLM $B - L$ parameter space, best excluded by the fifth-force satellite experiment MICROSCOPE \cite{MICROSCOPE:2022doy} and torsion-balance experiment E\"ot-Wash \cite{Wagner:2012ui}.  This showcases the power of such sensors in searching for this DM candidate, as was first pointed out in Ref.~\cite{Graham:2015ifn}. In a single-peak analysis focusing solely on the Compton peak, our limits would be weakened by approximately a factor of $2$, as can be seen from \cref{fig:beta-lim}. This difference becomes more dramatic for sensors located closer to the equator.

For the existing limits outlined above, we extract the E\"ot-Wash  limit from \cite{AxionLimits}; however, we recompute the MICROSCOPE limit. This is because the result of \cite{AxionLimits}, at the time of writing, is based on \cite{Berge:2017ovy}, which computed constraints based on the first MICROSCOPE results \cite{Touboul:2017grn,MICROSCOPE:2019jix}. We update this limit to reflect the final results given in \cite{MICROSCOPE:2022doy} following the reasoning outlined in \cite{Fayet:2017pdp, Fayet:2018cjy}. See \cref{app:microscope} for details. At low masses, we find that $g_{B - L} \lesssim 7 \times 10^{-26}$, improving the limit given in \cite{AxionLimits}, which we also show in \cref{fig:b-l}, by approximately a factor of $6.2$. For the $B - L$ limits computed in \cite{Fayet:2017pdp,Fayet:2018cjy}, we find an improvement by approximately a factor of $2.6$.

We only extend our limits to where they are appropriate. At the higher mass range, we are limited by keeping our observation time shorter than the coherence time, $T_\mathrm{obs} \ll \tau_\mathrm{coh} \simeq \SI{10}{\day}\,(5\times10^{-15}\,{\rm eV}/m)$. At the lower mass range, we must keep our observation time longer than a day (corresponding to $m \simeq 5\times10^{-20}$ eV) to be able to resolve the three peaks $T_\mathrm{obs} \gtrsim 2 \pi / \omega_\oplus$. This leads to the mass range $\SI{5e-20}{\electronvolt} \lesssim m \lesssim \SI{5e-15}{\electronvolt}$ for where our study is valid.

We note that $\sigma(\omega)$, though mostly slowly varying in the frequency width $\Delta \omega = \omega_\oplus$, does exhibit a large gradient around the resonance frequency of the cavity. In the neighbourhood of this frequency, our assumption that $\sigma(\omega)$ does not vary greatly in the above range is incorrect, and a more careful analysis in which all three peaks take on different noise levels would have to be conducted for more representative limits. However, we do not expect that our limits would differ greatly from our calculation and, at any rate, would still smoothly join to the regimes on either side of the resonance frequency, where our assumption holds.

\section{Future Directions}
\subsection{Longer Observation Times} In this work, we have focused on the wave vector DM signal within time scales much shorter than the coherence time, $\tau_{\rm coh} = 2 \pi /mv_0^2 \simeq \SI{50}{\day}\,(10^{-15}\,{\rm eV}/m)$. This allowed us to treat the amplitudes and phases of the three different components of the vector to be constant random variables. Realistically, there would be modulations giving these random variables time dependence---corrections of the order $\mathcal{O}(t/\tau_{\rm coh})$. Simultaneously, there would be spatial variations/correlations due to the finite distance covered by the Earth/detector during the observation, leading to corrections of the order $\mathcal{O}(v_0t/\ell_{\rm coh}) = \mathcal{O}(t/\tau_{\rm coh})$ again. For longer observation times, $T_{\rm obs} \gg \tau_{\rm coh}$, such modulations necessarily need to be taken into account. While we can easily generate the realistic time-series for such long time-scale signals (see top panel of \cref{fig:signals}), a more comprehensive treatment of how it affects our limits is left for future work.

Nevertheless, we comment on a simplified study that could be done within the \textit{incoherent} regime. In this limit, the stochasticity of the field is averaged over as $\mathcal{O}(T_\mathrm{obs}/\tau_\mathrm{coh})$ coherent patches cross the Earth. Thus, for $T_\mathrm{obs} \gg \tau_\mathrm{coh}$, the randomness in the signal disappears, and we are left with a deterministic signal. Moreover, the signal in Fourier space loses its coherent $T_\mathrm{obs}$ enhancement, reaching its maximal value at $T_\mathrm{obs} = \tau_\mathrm{coh}$ (c.f.~\cref{eq:beta}).

In a simplified experimental study, one could then analyse the data by first splitting the long-time time series into $N_\mathrm{coh} \sim T_\mathrm{obs} / \tau_\mathrm{coh}$ smaller, independent time series of coherence-time durations. Each one of these series would lead to our three-peak signal in Fourier space with randomly drawn Rayleigh amplitudes and uniform phases. One could then average over all $N_\mathrm{coh}$ PSDs, resulting in a deterministic amplitude where the randomness is no longer manifest. Crucially, this procedure would lead to the noise within each signal-containing bin to also be averaged over, resulting in a noise suppression by the factor $N_\mathrm{coh}^{-1/2}$. Similar arguments have been made in Refs.~\cite{Carney:2019cio,Manley:2020mjq}. To make inferences, we could then proceed by redefining our $\beta$ parameter as an `incoherent' version of it,
\begin{equation}
    \beta \rightarrow \beta_\mathrm{inc} \equiv \sqrt{\frac{\mathcal{A}^2}{2 \sigma^2}}(\tau_\mathrm{coh} T_\mathrm{obs})^{1/4}\,.
\end{equation}
This parameter can then be used in the deterministic likelihood, which is of the form of a non-central $\chi^2$ distribution (c.f.~\cref{eq:lik-full}), to find the limit $\beta_\mathrm{incoh}^\mathrm{lim}$ given the observed (averaged) PSD.

\subsection{Expanding the Mass Window}
\label{sec:expand_mass_window}
The detection scheme considered in this paper is one where the detector points in a specific fixed direction locally on Earth while rotating along with it. As noted above, in the short observation time scenario we are bounded from below by a day (corresponding to $m \simeq 5\times 10^{-20}$ eV) and from above by the coherence time scale $\tau_{\rm coh} \simeq \SI{10}{\day}\,(5\times10^{-15}\,{\rm eV}/m)$ to be able to resolve the distinctive $3$ peaks. To have a larger working window where we can neglect the previously mentioned corrections, the masses we can probe using this setup lie in the range $5\times 10^{-20}\,{\rm eV} \lesssim m \lesssim 5\times10^{-15}\,{\rm eV}$. To expand this window towards higher masses, we need to push the lower bound coming from $T_{\rm obs} > \SI{1}{\day}$. 

Instead of a fixed detector rotating with the Earth, a detector can also be made to rotate at an angular frequency faster than a day. For example, if  $\omega_{\rm exp} \simeq 2\pi/(1\,{\rm min})$, then the detector needs to collect data for only a few minutes to be able to isolate the three peaks. With $\tau_{\rm coh} \simeq 15\,{\rm min}\,(5\times 10^{-12}\,{\rm eV}/m)$, we can probe masses up to $m \simeq 5\times 10^{-12}$ eV. Note that while the time signal would contain modulations due to Earth's rotation, the effects would be suppressed by $\mathcal{O}(\omega_{\oplus}/\omega_{\rm exp})$. Furthermore, in the frequency space, this modulation would split the three peaks at $\omega_{\rm left} = m-\omega_{\rm exp}$, $\omega_{\rm middle} = m$, and $\omega_{\rm right} = m+\omega_{\rm exp}$, into nine peaks (each splitting into three). However, unless the frequency resolution $\Delta\omega$ becomes smaller than $\sim \omega_{\oplus}$, the experiment would not be able to resolve this splitting due to the Earth's rotation. Thus, we expect our whole analysis in this paper to carry forward, with $\omega_{\oplus}$ replaced by $\omega_{\rm exp}$ everywhere. Furthermore, such a setup would also be beneficial for the optomechanical light cavities we considered in \cref{sec:application}, where their typical laser coherence times are of the order of hours and not days.

\section{Conclusions}
\label{sec:concls}

We have provided an analysis strategy for inferring the properties of ultralight vector dark matter from terrestrial experiments, taking into account the stochastic and vector nature of the field (see \cref{fig:Ellipse}). Our main results are suited for observation times that are longer than a sidereal day, but shorter than the coherence time. They are as follows:
\begin{itemize}
    \item We focused on the signal in Fourier space, deriving the power spectral density that such dark matter is expected to leave on an axial sensor that is sensitive to its oscillatory signal. Accounting for the rotation of the Earth, we found that the signal manifests as three peaks at definite frequencies but with random amplitudes (see \cref{fig:signals}). 

    \item We derived the likelihoods in each of the signal-containing bins in Fourier space. We did this by considering the marginal likelihood after integrating out the six random variables exhibited by the ULDM signal in the coherent regime: the three Rayleigh amplitudes and the three uniformly distributed DM phases (see \cref{eq:liks} and \cref{fig:marginal-lik}). We found that the general elliptical motion of the vector field, arising out of equipartition, afforded us a simpler analysis in Fourier space than the linear polarization assumption. This is because in the former, all peaks become statistically uncorrelated.

    \item We drew exclusion limits on a generalised, dimensionless parameter that can be re-interpreted in the context of a concrete sensor setup and dark matter model. We did this via a series of log-likelihood ratio tests following a hybrid frequentist-Bayesian approach. Crucially, we found that, unlike analyses focusing on only a single peak, our approach retains constraining power for experimental setups at all latitudes. This is because we make use of the entire DM signal, which is distributed across all three peaks, instead of constraining ourselves to the signal in any one peak, which is dependent on the latitude of the experiment (see \cref{fig:beta-lim}).

    \item We considered a specific sensor technology (the optomechanical light cavity) and dark matter model (ultralight dark matter stemming from a new gauged $U(1)_{B - L}$ symmetry) as a concrete application of our analysis strategy. We recast our general limit onto one on the gauge coupling of this model, $g_{B-L}$, finding that long-exposure cavities can rule out previously unexplored regions of the $B - L$ parameter space (see \cref{fig:b-l}). 
\end{itemize}

In this work, we have established a framework for future experimental efforts in the detection of ultralight vector dark matter. Novel direct-detection probes require an understanding of how the signal of ultralight vector dark matter behaves in our local neighborhood and manifests itself in a sensor.  We hope that our work aids in (i) designing search strategies using emerging detector technologies that are not traditionally used for dark matter searches, and (ii) in understanding how well a given model can be tested in the context of calls for Big Science projects using quantum sensing~\cite{Windchime:2022whs}.

\acknowledgments

We would like to thank Andrew Long for insightful discussions throughout this work, as well as John Carlton, Daniel Carney, Pierre Fayet, Zhen Liu, and Yue Zhao for their valuable feedback. We also thank Pierre Fayet for their comments regarding our MICROSCOPE limit  . We are grateful to Benjamin Schussler for creating a dynamical visualization  of the dark matter vector field in our halo (link to video in caption of Fig.~1). DA would like to thank Yunan Gao and Juehang Qin for helpful discussions regarding our statistical treatment. CT \& DA
are supported by the National Science Foundation
under award 2046549. MA \& MJ were partly supported by DOE grant DE-SC0021619, and MJ is currently supported by a Leverhulme Trust Research Project (RPG-2022-145).

\bibliographystyle{JHEP}
\bibliography{biblio}

\newpage

\appendix

\section{On the Longitudinal and Transverse Modes of the Vector Field}
\label{app:long_trans_modes_VDM}

\subsection{Signal for Arbitrary Power Spectrum}

In this section, we describe the stochastic behavior of the random vector dark matter field for the case that there is an unequal partitioning of power amongst its longitudinal and transverse modes.
Requiring the two-point correlation 
$\langle\hat{\bPsi}^{\dagger}_{\bm k}\cdot\hat{\bPsi}_{\bm p}\rangle = \delta_{\bm k,\, \bm p}\,f_{\bm k}$ (where $f_{\bvec{k}}$ is the halo function), we can parameterize the power spectrum as
\begin{align}
    \label{eq:Breaking_powerspectrum}    \langle\hat{\psi}^{\,i\,\ast}_{\bm k}\,\hat{\psi}^{\,j}_{\bm p}\rangle = \delta_{\bm k,\, \bm p}\,f_{\bm k}\left[\Upsilon_0\,\mathcal{P}^{ij}_{\bm k_\perp} + (1-2\Upsilon_0)\mathcal{P}^{ij}_{\bm k_{||}}\right]\,,
\end{align}
where
\begin{align}
    \mathcal{P}^{ij}_{\bm k_\perp} \equiv \left(\delta^{ij}-\frac{k^ik^j}{{|\bm k}|^2}\right)\qquad{\rm and}\qquad \mathcal{P}^{ij}_{\bm k_{||}} \equiv \frac{k^ik^j}{{|\bm k}|^2}
\end{align}
are the two mutually orthogonal projection matrices. The parameter $\Upsilon(t)$ quantifies the partitioning of the power spectrum into longitudinal and transverse modes at arbitrary times. In particular, for our purposes here, $\Upsilon_0 \equiv \Upsilon(t=t_0)$ is the partitioning in the halo today. While $\Upsilon \in [0,1/2]$ in general, $\Upsilon=0$ gives a longitudinal-only spectrum, and $\Upsilon=1/2$ gives a transverse-only spectrum. For $\Upsilon = 1/3$, we achieve equipartition. In greater generality, we note that $\Upsilon$ could be $\bvec{k}$-dependent; we leave a detailed analysis of this case for future work.

Rescaling $\hat{\bPsi}_{\bm k}$ such that $\hat{\bPsi}_{\bm k}=\sqrt{f_{\bm k}}\,\hat{\bepsilon}_{\bm k}$ gives the two-point correlation as $\langle\hat{\epsilon}^{\,i\,\ast}_{\bm k}\,\hat{\epsilon}^{\,j}_{\bm p}\rangle = \delta_{\bm k,\, \bm p}\left[\Upsilon_0\,\mathcal{P}^{ij}_{\bm k_\perp} + (1-2\Upsilon_0)\mathcal{P}^{ij}_{\bm k_{||}}\right]$. We can break $\bepsilon$ into a transverse piece, $\hat{\epsilon}^{\,i}_{\bm k_{\perp}} = \mathcal{P}^{ij}_{\bm k_{\perp}}\,\hat{\epsilon}^{\,j}_{\bm k}$, and a longitudinal piece, $\hat{\epsilon}^{\,i}_{\bm k_{||}} = \mathcal{P}^{ij}_{\bm k_{||}}\,\hat{\epsilon}^{\,j}_{\bm k}$, with zero cross-correlation and following self-correlations:
\begin{align}
    \langle\hat{\epsilon}^{\,i\,\ast}_{\bm k_{\perp}}\,\hat{\epsilon}^{\,j}_{\bm p_{\perp}}\rangle = \Upsilon_0\,\delta_{\bm k,\,\bm p}\,\mathcal{P}_{\bm k_{\perp}}^{\,ij}\qquad \text{and} \qquad \langle\hat{\epsilon}^{\,i\,\ast}_{\bm k_{||}}\,\hat{\epsilon}^{\,j}_{\bm p_{||}}\rangle = (1-2\Upsilon_0)\delta_{\bm k,\, \bm p}\,\mathcal{P}_{\bm k_{||}}^{\,ij}\,.
\end{align}
The above suggests that, for every ${\bm k}$, we can pick a real $6$-dimensional random variable $\hat{\bm u}_{\bm k}$ restricted to lie uniformly on an $S_5$. We can package this as a $3$-dimensional complex random variable (i.e. $\langle\hat{u}^{\,i\,\ast}_{\bm k}\,\hat{u}^{\,j}_{\bm p}\rangle = \delta_{\bm k,\, \bm p}\,\delta^{ij}$), and then hit it with the transverse and longitudinal projection operators scaled by $\sqrt{\Upsilon_0}$ and $\sqrt{1-2\Upsilon_0}$, respectively, to get $\hat{\bepsilon}_{\bm k_{\perp}}$ and $\hat{\bepsilon}_{\bm k_{||}}$.\\

With this, the full field can be broken into longitudinal and transverse pieces:
\begin{align}
    \hat{A}^i({\bm x},\,t) = \sqrt{\frac{2}{m}}\,\Re\Biggl\{\frac{1}{\sqrt{V}}\sum_{\bm k}\,e^{i{\bm k}\cdot{\bm x}}\,\sqrt{f_{\bm k}}\,\left(\sqrt{\Upsilon_0}\,\mathcal{P}_{\bm k_{\perp}}^{\,ij} + \sqrt{1-2\Upsilon_0}\,\mathcal{P}_{\bm k_{||}}^{\,ij}\right)\hat{u}^{\,j}_{\bm k}\,e^{-i(m + \frac{k^2}{2m})t}\Biggr\}\,.
\end{align}
Using the central limit theorem, this yields the following:
\begin{align}
    \hat{A}^j({\bm 0},\,t)\Bigr|_{t \ll T_{\rm coh}} = \sqrt{\frac{2}{m}}\Re\left[e^{-imt}(\sqrt{\Upsilon_0}\,\hat{a}^j + \sqrt{1-2\Upsilon_0}\,\hat{b}^j)\right]\,,
\end{align}
where $\hat{\bm a}$ and $\hat{\bm b}$ are two $6$-dimensional independent random variables (represented as two $3$-dimensional complex random variables) with zero mean and following self-correlations in the large volume limit:
\begin{align}
\label{eq:P_squared_gamma}
    \langle \hat{a}^{i\ast}\,\hat{a}^j\rangle &= \int\frac{\mathrm{d}^3\bvec{k}}{(2\pi)^3}f_{\bm k}\,\mathcal{P}^{ij}_{\bm k_{\perp}} = \frac{\rho}{m}\left[(1-\mathcal{F})\delta^{ij} - (1-3\mathcal{F})\frac{\bar{k}^i\bar{k}^j}{\bar{k}^2}\right]\nonumber\\
    \langle\hat{b}^{i\ast}\,\hat{b}^j\rangle &= \int\frac{\mathrm{d}^3\bvec{k}}{(2\pi)^3}f_{\bm k}\,\mathcal{P}^{ij}_{\bm k_{||}} = \frac{\rho}{m}\left[\mathcal{F}\,\delta^{ij} + (1-3\mathcal{F})\frac{\bar{k}^i\bar{k}^j}{\bar{k}^2}\right]\,.
\end{align}
Here, 
\begin{align}
    \mathcal{F} = \frac{(m\sigma)^2}{\bar{k}^2}\left[1 - \sqrt{\frac{\pi}{2}}\frac{m\sigma}{\bar{k}}e^{-\frac{\bar{k}^2}{2(m\sigma)^2}}{\rm Erfi}\left(\frac{\bar{k}}{\sqrt{2}m\sigma}\right)\right]\,,
\end{align}
and the cross correlation between $\hat{\bm a}$ and $\hat{\bm b}$ is zero, i.e. $\langle\hat{a}^{i\,\ast}\hat{b}^j\rangle = 0$. Here, $\bar{k} / m$ is our velocity relative to the rest frame of the halo, and $\sigma$ is the velocity dispersion. Owing to their Gaussian nature, we can further combine the two random variables and define $\hat{w}^j \equiv \Upsilon_0\,\hat{a}^j + \sqrt{1-2\Upsilon_0}\,\hat{b}^j$. This new combined random variable will have zero mean, and its variance will be equal to the sum of the individual two in Eq.~\eqref{eq:P_squared_gamma} (weighted by $\Upsilon_0$ and $(1-2\Upsilon_0)$, respectively). Also, we can extract the factor $\sqrt{\rho/m}$. With all of this, we have that
\begin{align}
\label{eq:A_decompose}
    \hat{A}^j({\bm 0},\,t)\Bigr|_{t \ll T_{\rm coh}} = \frac{\sqrt{2\rho}}{m}\Re\left[e^{-imt}\,\hat{w}^j\right]\,,
\end{align}
where $\langle\hat{w}^j\rangle = 0$ and 
\begin{align}
\label{eq:cc_corr}
    \langle \hat{w}^{i\ast}\,\hat{w}^j\rangle = \mathcal{W}^{ij} \equiv \left[(\mathcal{F} - 3\Upsilon_0 \mathcal{F} + \Upsilon_0)\,\delta^{ij} + (1-3\mathcal{F})(1-3\Upsilon_0)\frac{\bar{k}^i\bar{k}^j}{\bar{k}^2}\right]\,.
\end{align}
With the form of the stochastic vector field derived, we now need to generate these random variables $\hat{\bm w}$s. We can do this by finding an operator matrix $\mathcal{G}$ such that its square gives the right-hand side of the above two-point correlation in \cref{eq:cc_corr}. Then, we can simply pick a $3$-dimensional normal complex random variable, say $\hat{\bm h}$ (which can be equivalently thought of as a $6$-dimensional normal random variable), and hit it with $\mathcal{G}$ to get $\hat{\bm w}$. That is, we have
\begin{align}
    \hat{w}^i &= \mathcal{G}^{ij}\hat{h}^j\,,
\end{align}
where
\begin{align}
    \mathcal{G}^{ij} &= \sqrt{\mathcal{F}-3\Upsilon_0 \mathcal{F}+\Upsilon_0}\left[\delta^{ij} - \left(1 - \sqrt{\frac{1 - 2(\mathcal{F}-3\Upsilon_0 \mathcal{F}+\Upsilon_0)}{\mathcal{F}-3\Upsilon_0 \mathcal{F}+\Upsilon_0}}\right)\frac{\bar{k}^i\bar{k}^j}{\bar{k}^2}\right]\,.
\end{align}
While this serves as a procedure to generate the stochastic random vector field $\hat{\bm A}$, the problem simplifies dramatically when the power is equipartitioned between the longitudinal and transverse modes. That is, when $\Upsilon_0 = 1/3$. We expect this to be the case when the vector field accounts for (at least the majority of) the virialized dark matter around us. We discuss this next. 

\subsection{Equipartition between Longitudinal and Transverse Modes}
\label{app:equipartition}

The equation of motion for the field $\bPsi$ (written in Fourier space) is
\begin{align}
\label{eq:Schreqn_kspace}
    i\dot{\Psi}^i_{\bm k} &= \frac{{|\bm k|}^2}{2m}\Psi^i_{\bm k} - (4\pi Gm^2)\frac{1}{V}\sum_{\bm p,\, \bm q,\, \bm \ell}\delta_{\bm k + \bm p - \bm q - \bm \ell}\,|{\bm q}-{\bm p}|^{-2}\,\Psi^{j\,\ast}_{\bm p}\Psi^j_{\bm q}\Psi^i_{\bm \ell}\,,
\end{align}
where $E_{\bm k} = {|\bm k|}^2/2m$. Now, we wish to decompose the field into transverse and longitudinal pieces. Let us call them $\eta_{\bm k}$ and $\zeta_{\bm k}$. That is,
\begin{align}
    \eta_{\bm k}^i = \mathcal{P}^{ij}_{\bm k_\perp}\,\Psi^j_{\bm k}\qquad{\rm and}\qquad \zeta_{\bm k}^i = \mathcal{P}^{ij}_{\bm k_{||}}\,\Psi^j_{\bm k}\,,
\end{align}
where $\mathcal{P}^{ij}_{\bm k_\perp}$ and $\mathcal{P}^{ij}_{\bm k_{||}}$ are the two mutually orthogonal projection matrices defined in the previous section. Let us hit the above equation of motion by the transverse (longitudinal) projection operator to get the evolution for $\eta^i_{\bm k}$ ($\zeta^i_{\bm k}$). Also splitting each $\Psi^i_{\bm k}$ into $\eta^i_{\bm k}$ and $\zeta^i_{\bm k}$, we get
\begin{align}
    i\dot{\eta}_{\bm k}^i &= E_{\bm k}\eta_{\bm k}^i - (4\pi G m^2)\frac{1}{V}\sum_{\bm p,\, \bm q,\, \bm \ell}\delta_{\bm k + \bm p - \bm q - \bm \ell}\,|{\bm q}-{\bm p}|^{-2}\,(\eta^{n\,\ast}_{\bm p}+\zeta^{n\,\ast}_{\bm p})(\eta^{n}_{\bm q}+\zeta^{n}_{\bm q})(\mathcal{P}^{ij}_{\bm k_\perp}\eta^j_{\bm \ell} + \mathcal{P}^{ij}_{\bm k_\perp}\zeta^j_{\bm \ell})\nonumber\\
    i\dot{\zeta}_{\bm k}^i &= E_{\bm k}\zeta_{\bm k}^i - (4\pi G m^2)\frac{1}{V}\sum_{\bm p,\, \bm q,\, \bm \ell}\delta_{\bm k + \bm p - \bm q - \bm \ell}\,|{\bm q}-{\bm p}|^{-2}\,(\eta^{n\,\ast}_{\bm p}+\zeta^{n\,\ast}_{\bm p})(\eta^{n}_{\bm q}+\zeta^{n}_{\bm q})(\mathcal{P}^{ij}_{\bm k_{||}}\eta^j_{\bm \ell} + \mathcal{P}^{ij}_{\bm k_{||}}\zeta^j_{\bm \ell})\,.
\end{align}
Notice that not only are the longitudinal and transverse pieces coupled to one another, but more importantly they also source each other. Hence, we expect virialization between them due to such non-linear gravitational dynamics during DM halo formation, ultimately leading to the equipartition of power amongst them. That is, $2/3$ of the total power would be in the transverse sector, with the remaining $1/3$ in the longitudinal sector.\\

Using codes based on the split-Fourier technique~\cite{Mocz:2017wlg,Edwards:2018ccc,Jain:2022agt,Jain:2023qty}, we have performed simulations of the \schr-Poisson vector system~\cite{Adshead:2021kvl,Jain:2021pnk,Amin:2022pzv} (physical space version of \cref{eq:Schreqn_kspace}):
\begin{align}
\label{eq:SP_system}
    i\frac{\partial}{\partial t}\bPsi &= -\frac{1}{2m}\nabla^2\bPsi + m\,\Phi\,\bPsi\,,\qquad\qquad
    \nabla^2\Phi = 4\pi Gm^2\,\bPsi^\dagger \bPsi\,,
\end{align}
with different initial conditions. Parameterizing the initial power spectrum by $\Upsilon(0)$, c.f.~\cref{eq:Breaking_powerspectrum}, we have
\begin{align}
    \langle\hat{\psi}^{\,i\,\ast}_{\bm k}\,\hat{\psi}^{\,j}_{\bm p}\rangle = \delta_{\bm k,\, \bm p}\,\tilde{f}_{\bm k}\left[\Upsilon(0)\,\mathcal{P}^{ij}_{\bm k_\perp} + (1-2\Upsilon(0))\mathcal{P}^{ij}_{\bm k_{||}}\right]\,,
\end{align}
where $\tilde{f}_{\bm k}$ is some arbitrary Fourier profile function (we shall take it to be a Gaussian in our simulations, such that the mean density is larger than the critical Jeans density so as to form a halo). 

\begin{figure}
    \includegraphics{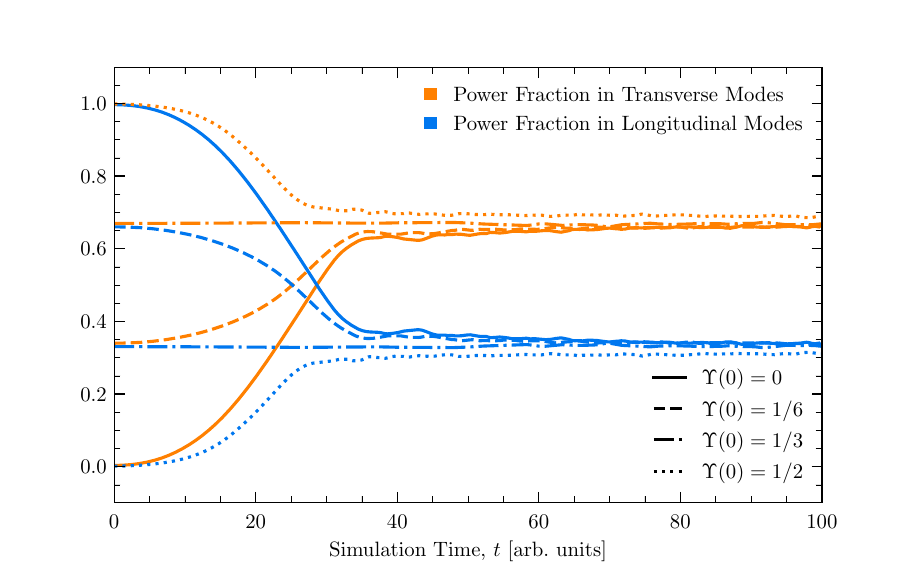}
    \hfill
    \centering
    \caption{Evolution of the fractional powers in the transverse and longitudinal modes (Eq.~\eqref{eq:fraction_powers}) around halo formation for four different simulations with different $\Upsilon(0)$: $\Upsilon(0) = 0$ (solid), ${\Upsilon(0) = 1/6}$ (dashed), $\Upsilon(0) = 1/3$ (dot-dashed), and  $\Upsilon(0) = 1/2$ (dotted). The convergence of the fractional power in transverse modes towards $2/3$ and that in longitudinal modes towards $1/3$ demonstrates the ultimate equipartition of power. That is, $\Upsilon(t) \rightarrow 1/3$.}
    \label{fig:powerfraction}
\end{figure}

To begin with, for every mode ${\bm k}$, we pick a real $6$-dimensional normal random variable, packaged as a $3$-dimensional complex number $\hat{\bepsilon}_{\bm k}$, that is further restricted to lie uniformly on an $S_5$ so that it is unit normalized. That is, $\langle\hat{\epsilon}^{i\ast}_{\bm k}\hat{\epsilon}^j_{\bm p}\rangle = \delta_{\bm k,\, \bm p}\,\delta^{ij}$. We then hit it with the transverse and longitudinal projection operators to get $\hat{\epsilon}^i_{\bm k_{\perp}} = \mathcal{P}^{ij}_{\bm k_{\perp}}\hat{\epsilon}^j_{\bm k}$ and $\hat{\epsilon}_{\bm k_{||}} = \mathcal{P}^{ij}_{\bm k_{||}}\hat{\epsilon}^j_{\bm k}$. With these, we construct the full transverse and longitudinal initial fields as
\begin{align}
    \{\eta^i(\bm x,\,\, 0),\zeta^i(\bm x,\, 0)\} = \frac{1}{\sqrt{V}}\sum_{\bm k}\{\eta^{i}_{\bm k}(0),\, \zeta^{i}_{\bm k}(0)\}\,e^{-i{\bm k}\cdot{\bm x}} = \frac{1}{\sqrt{V}}\sum_{\bm k}\sqrt{\tilde{f}_{\bm k}}\,\{\hat{\epsilon}^i_{\bm k_{\perp}},\, \hat{\epsilon}^i_{\bm k_{||}}\}\,e^{-i{\bm k}\cdot{\bm x}}\,,
\end{align}
with the full field being 
\begin{align}
    \Psi^i({\bm x},\,0) = \sqrt{\Upsilon(0)}\,\eta^i({\bm x},\,0) + \sqrt{1-2\Upsilon(0)}\,\zeta^i({\bm x},\,0)\,.
\end{align}
With this as the initial condition, we evolve the SP system~\eqref{eq:SP_system}. In~\cref{fig:powerfraction}, we present simulation results for various different initial conditions, including different values of $\Upsilon(0)$. Here, we plot the fractional powers in the two different sectors:
\begin{align}
\label{eq:fraction_powers}
    F_{\perp}(t) = \frac{\int{\rm d}^3 \bvec{x}\,[{\bm \eta}^{\ast}({\bm x},\,t)\cdot{\bm \eta}({\bm x},\,t)]}{\int{\rm d}^3\bvec{x}\,[{\bm \psi}^{\ast}({\bm x},\,t)\cdot{\bm \psi}({\bm x},\,t)]}\quad{\rm and}\quad F_{||}(t) = \frac{\int{\rm d}^3 \bvec{x}\,[{\bm \zeta^{\ast}({\bm x},\,t)}\cdot{\bm \zeta({\bm x},\,t)}]}{\int{\rm d}^3\,\bvec{x}\,[{\bm \psi^{\ast}({\bm x},\,t)}\cdot{\bm \psi({\bm x},\,t)}]}\,.
\end{align}
With $\Upsilon \rightarrow \Upsilon_0 = 1/3$, we achieve equipartition. The stochastic vector field, c.f.~\cref{eq:A_decompose}, takes the simple form
\begin{align}
    \hat{A}^j({\bm 0},\,t)\Bigr|_{t \ll T_{\rm coh}} = \frac{\sqrt{2\rho}}{m}\Re\left[e^{-imt}\,\hat{w}^j\right],\quad{\rm where}\quad \langle\hat{w}^j\rangle = 0\quad{\rm and}\quad\langle \hat{w}^{i\ast}\,\hat{w}^j\rangle = \frac{1}{3}\delta_{ij}\,,
\end{align}
and is the form that we have used in~\cref{eq:A_decompose_main}. This gives rise to the ``random ellipse" picture, discussed in the main text.

To conclude, we have shown that even if the vector field is initialized with an unequal distribution of power between its longitudinal and transverse modes, non-linear gravitational dynamics leads to its equipartition eventually. However, we have only begun to explore this topic and leave a detailed study (including a $\bvec{k}$-dependent $\Upsilon$) for future work.

\section{Derivation of Marginal Likelihood with Stochastic Field Amplitude}
\label{app:deriv-marginal}

The full signal in time space is given by
\begin{equation}
    \hat{\mathcal{S}}(t) = \mathrm{signal}(t) + \hat{\mathcal{N}}(t)\,,
\end{equation}
where $\hat{\mathcal{N}} \sim N(0, \sigma_t)$, with $\sigma_t^2 \equiv \sigma^2 / \Delta t$ and $\sigma^2$ being the equivalent of the noise PSD in frequency space.
From this, we get that the (two-sided) periodogram, $\mathcal{P}'$, normalised by the noise PSD is given by
\begin{equation}
    \frac{\hat{\mathcal{P}}'}{\sigma^2} = \frac{\Delta t^2}{T}\frac{|\hat{\mathcal{S}}^2(\omega)|} {\sigma^2} = |\hat{X}_R + {\rm signal}_R(\omega)|^2 + |\hat{X}_I + {\rm signal}_I(\omega)|^2\,.
\end{equation}
Here, both $\hat{X}_R$ and $\hat{X}_I$ are such that $\hat{X}_i \sim N(0, 1 / \sqrt{2})$. The one-sided, noise-normalised periodogram, $\hat{\mathcal{P}} / \sigma^2$, therefore follows a non-central $\chi^2$ distribution with non-centrality parameter
\begin{align}
\label{eq:lambda}
    \hat{\lambda} &= {\rm signal}^2_R(\omega) + {\rm signal}^2_I(\omega)\nonumber\\
    &= \frac{\mathcal{A}^2T}{8\,\sigma^2}\Bigl\{\cos^2\phi\left[\hat{\alpha}_x^2 + \hat{\alpha}_y^2 + 2 \hat{\alpha}_x \hat{\alpha}_y \sin(\hat{\varphi}_y - \hat{\varphi}_x)\right]\delta_{\omega,s}\nonumber\\
    &\qquad\quad + \cos^2\phi\left[\hat{\alpha}_x^2 + \hat{\alpha}_y^2 - 2 \alpha_x \hat{\alpha}_y \sin(\hat{\varphi}_y -\hat{\varphi}_x)\right]\delta_{\omega,d}\nonumber\\
    &\qquad\quad + 4 \sin^2\phi\,\hat{\alpha}_z^2 \delta_{\omega,m}\Bigr\}\,.
\end{align}
Introducing randomness in the parameters $\alpha$'s and $\varphi$'s, with prior $\Pi'(\{\alpha_i,\varphi_i\})$, the marginalized likelihood is
\begin{align}
    \mathcal{L} &= \int\mathrm{d}\alpha_x\mathrm{d}\alpha_y\mathrm{d}\alpha_z\mathrm{d}\varphi_x\mathrm{d}\varphi_y\,\Pi'(\varphi_x,\varphi_y,\alpha_x,\alpha_y,\alpha_z)\,e^{-(\lambda + p)/2}I_0(\sqrt{\lambda p})\,,
\end{align}
where $I_0$ is the modified Bessel function of the first kind. To evaluate the above integral, we first note that the prior is factorizable into that for $\alpha_z$ and for the set $\{\alpha_x,\alpha_y,\varphi_x,\varphi_y\}$. The latter $4$ random variables (which correspond to the $\omega = s$ and $\omega = d$ peaks), can be redefined as two $2$D random vectors $\bvec{x}$ and $\bvec{y}$ with relative angle $\pi/2 - (\varphi_x - \varphi_y)$, in order to give
$\alpha_x^2 + \alpha_y^2 \pm 2 \alpha_x \alpha_y \sin(\varphi_y - \varphi_x) \rightarrow |\bvec{x} \pm \bvec{y}|^2 = (x_1 \pm y_1)^2 + (x_2 \pm y_2)^2$. Here the subscript $1$ and $2$ correspond to the two components of the vectors, in the two directions of the 2D Euclidean space respectively. Since $\alpha$'s are Rayleigh distributed and $\varphi$'s are uniformly distributed from $(0,2\pi)$, the four variables $\{x_1,x_2,y_1,y_2\}$ are normally distributed with zero mean and variance equal to $1/2$. Furthermore, we can now redefine $x$'s and $y$'s as $x_i + y_i = u_i$ and $x_i - y_i = v_i$ for $i = \{1,2\}$, to get the following expression
\begin{align}
    \mathcal{L} &\propto \int\mathrm{d}\alpha_z\,\Pi'(\alpha_z)\mathrm{d}u_1\mathrm{d}u_2\mathrm{d}v_1\mathrm{d}v_2\,\Pi(u_1,u_2,v_1,v_2)\,e^{-(\lambda+p)/2}I_0(\sqrt{\lambda p})\,.
\end{align}
with $\Pi(u_1,u_2,v_1,v_2) = e^{-(u_1^2+u_2^2+v_1^2+v_2^2)/2}$, $\Pi'(\alpha_z) = 2\alpha_z\,e^{-\alpha_z^2}$, and
\begin{align}
    \lambda &= \frac{\mathcal{A}^2T}{8\,\sigma^2}\Bigl\{\cos^2\phi\left[u^2_1 + u^2_2\right]\delta_{\omega,s} + \cos^2\phi\left[v^2_1 + v^2_2\right]\delta_{\omega,d} + 4 \sin^2\phi\,\alpha_z^2 \delta_{\omega,m}\Bigr\}\,.
\end{align}
Using the series representation of the Bessel function, together with Gamma function identities, the $5$ random variables can be integrated out analytically. We arrive at the following marginalized (and normalized) likelihood:
\begin{align}
    \mathcal{L} &= \frac{e^{-\frac{p}{2}\left(\frac{1}{1+X\delta_{\omega,s}}\right)} + e^{-\frac{p}{2}\left(\frac{1}{1+X\delta_{\omega,d}}\right)} + e^{-p\left(\frac{1}{2+Y\delta_{\omega,m}}\right)} - 2e^{-\frac{p}{2}}}{2(1+X\delta_{\omega,s})+2(1+X\delta_{\omega,d})+(2+Y\delta_{\omega,m})-4}\,,
\end{align}
where
\begin{align}
    X = \frac{\mathcal{A}^2T}{8\,\sigma^2}\cos^2\phi \qquad {\rm and} \qquad Y = \frac{\mathcal{A}^2T}{2\,\sigma^2}\sin^2\phi\,.
\end{align}
This likelihood can be split into three individual likelihoods for the sum/difference peaks and the Compton peak, as given in \cref{eq:liks}. The form of the likelihoods for the sum and difference peaks is equivalent.

To treat the total likelihood as the product of the individual likelihoods in each frequency bin, we must check that the covariance matrix is diagonal.
We will consider a signal-only analysis, discarding the noise, since the noise merely adds to the power and is uncorrelated between different frequency bins. We may write the values of the three peaks as
\begin{equation}
\begin{split}
\hat{\mathcal{P}}_1 &= \frac{\mathcal{A}^2 T_\mathrm{obs}}{8}\left[\hat{\alpha}_x^2 + \hat{\alpha}_y^2 + 2 \hat{\alpha}_x \hat{\alpha}_y \sin(\hat{\varphi}_y - \hat{\varphi}_x)\right]\cos^2\phi\,,\\
\hat{\mathcal{P}}_2 &= \frac{\mathcal{A}^2 T_\mathrm{obs}}{8}\left[\hat{\alpha}_x^2 + \hat{\alpha}_y^2 - 2 \hat{\alpha}_x \hat{\alpha}_y \sin(\hat{\varphi}_y - \hat{\varphi}_x)\right]\cos^2\phi\,,\\
\hat{\mathcal{P}}_3 &=  \frac{\mathcal{A}^2 T_\mathrm{obs}}{2} \alpha_z^2 \sin^2\phi\,.
\end{split}
\end{equation}
We wish to compute the quantity
\begin{equation}
\Sigma_{ij} \equiv \langle \hat{\mathcal{P}}_i \hat{\mathcal{P}}_j\rangle - \langle \hat{\mathcal{P}}_i \rangle \langle \hat{\mathcal{P}}_j \rangle\,.
\end{equation}
We can do this using the expression for the raw moments,
\begin{equation}
\langle \alpha^n\rangle = 2^{n/2} \sigma^n \Gamma\left(1 + \frac{n}{2}\right)\,,
\end{equation} 
where, for us, $\sigma = 1 / \sqrt{2}$. Aside from this, we need to know that
\begin{equation}
\begin{split}
\langle \sin(\hat{\varphi}_y - \hat{\varphi}_x)\rangle &= 0\,, \\
\langle \sin^2(\hat{\varphi}_y - \hat{\varphi}_x)\rangle &= \frac{1}{2}\,.
\end{split}
\end{equation}
We then get the diagonal covariance matrix
\begin{equation}
\bvec{\Sigma} = 
    \frac{\mathcal{A}^4 T_\mathrm{obs}^2}{16}
    \begin{pmatrix}
    \cos^4\phi & 0 & 0 \\
    0 & \cos^4\phi & 0 \\
    0 & 0 & 4 \sin^4\phi
    \end{pmatrix}\,.
\end{equation}
Crucially, we get that the covariance between peaks is $0$, allowing us to treat them as statistically independent and hence permitting us to express the total likelihood as the product of the individual likelihoods. 
%~~~~~~~~~~~~~~~~~~~~~~~~~~~~~~~~~~
\begin{figure}[t!]
    \centering
    \includegraphics{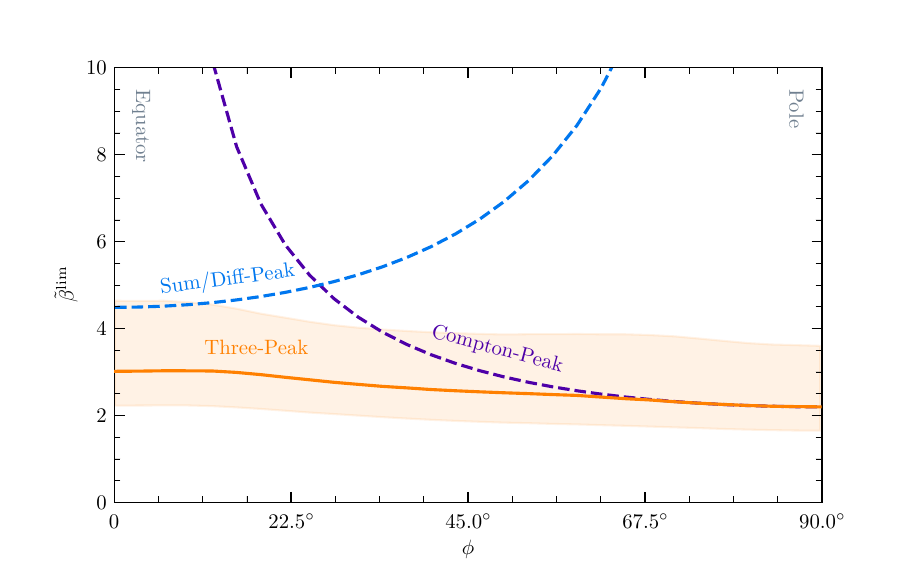}
    \caption{The $90\%$ CL limits ($\tilde{\beta}^\mathrm{lim}$) on the dimensionless parameter $\tilde{\beta} \equiv \sqrt{\rho\,g_{\rm eff}^2\,\sigma_v^2\,T_\mathrm{obs} / \sigma_n^2}$ (see \cref{eq:beta_grad_scalar}) derived from our MC analysis (the same as in \cref{fig:beta-lim} but for the case of the gradient of a scalar). We show the results of our three-peak analysis and those of two single-peak analyses focusing on the Compton peak and on one of the sum or difference peaks. The shaded region indicates the $1\sigma$ error bar on our three-peak analysis. Our three-peak analysis generally provides a better/comparable limit to either of the two single-peak analyses and is largely latitude-independent.}
    \label{fig:grad-scalar-beta}
\end{figure}
%~~~~~~~~~~~~~~~~~~~~~~~~~~~~~~~~~~
%%%%%%%%%%%%%%%%%%%%%%%%%%%%%%%%%%%%%%%%%%%%%%%
\section{The Case of the Gradient of a Scalar}
\label{app:likelihood_gradientscalar}
%%%%%%%%%%%%%%%%%%%%%%%%%%%%%%%%%%%%%%%%%%%%%%%

In this case, there is a preferential direction because $\bm{\nabla} a$ points in the direction of the local DM velocity. Aligning the lab's working coordinate system such that this local velocity vector is parallel to the $z$ axis, the amplitudes associated with the three different directions in~\cref{eq:low-t-period} are not all the same. Effectively, there is an extra factor associated with the $z$ direction, and the random signal in frequency space (c.f.~\cref{eq:low-t-period}) takes the following form
\begin{equation}
\label{eq:low-t-period_gos}
\begin{split}
\hat{\tilde{\lambda}}(\omega_n) = \frac{\tilde{\beta}^2}{4}\Big\{&\left[\hat{\alpha}_x^2 + \hat{\alpha}_y^2 + 2 \hat{\alpha}_x \hat{\alpha}_y \sin(\hat{\varphi}_y - \hat{\varphi}_x)\right]\cos^2\phi\,\delta_{\omega_n,s}\\ 
+&\left[\hat{\alpha}_x^2 + \hat{\alpha}_y^2 - 2 \hat{\alpha}_x \hat{\alpha}_y \sin(\hat{\varphi}_y -\hat{\varphi}_x)\right]\cos^2\phi\,\delta_{\omega_n,d}\\
+&~4 \tilde{\Upsilon}\,\hat{\alpha}_z^2 \sin^2\phi\,\delta_{\omega_n,m}\Big\}\,,
\end{split}
\end{equation}
where (and following the notation of~\cite{Lisanti:2021vij})
\begin{align}
    \label{eq:beta_grad_scalar}
  \tilde{\beta} \equiv \sqrt{\frac{\rho\,g_{\rm eff}^2\,\sigma_v^2\,T_\mathrm{obs}}{\sigma_n^2}}\qquad{\rm and}\qquad \tilde{\gamma} \equiv 1+\left(\frac{v_{\odot}}{\sigma_v}\right)^2 \simeq 2.1 \,.
\end{align}
Proceeding similarly as in \cref{app:deriv-marginal},
the marginalized likelihood is
\begin{align}
    \mathcal{L}' &= \int\mathrm{d}\alpha_x\mathrm{d}\alpha_y\mathrm{d}\alpha_z\mathrm{d}\varphi_x\mathrm{d}\varphi_y\,\Pi'(\varphi_x,\varphi_y,\alpha_x,\alpha_y,\alpha_z)\,e^{-(p+\tilde{\lambda})/2}I_0(\sqrt{p\tilde{\lambda}})\,,
\end{align}
which we can evaluate by proceeding in the same fashion as in \cref{app:deriv-marginal}; i.e. making redefinitions of the variables so they become independent and the integral becomes analytically tractable. We arrive at the following:
\begin{align}
    \mathcal{L}' &= \frac{e^{-\frac{p}{2}\left(\frac{1}{1+X\delta_{\omega,s}}\right)} + e^{-\frac{p}{2}\left(\frac{1}{1+X\delta_{\omega,d}}\right)} + e^{-\frac{p}{2}\left(\frac{1}{1+Y\delta_{\omega,m}}\right)} - 2e^{-\frac{p}{2}}}{2(1+X\delta_{\omega,s})+2(1+X\delta_{\omega,d})+2(1+Y\delta_{\omega,m})-4}\,,
\end{align}
where
\begin{align}
    X = \frac{\tilde{\beta}^2}{4}\cos^2\phi \qquad {\rm and} \qquad Y = \frac{\tilde{\beta}^2\tilde{\gamma}}{2}\sin^2\phi\,.
\end{align}
The overall result is that $Y$ simply gets rescaled by $\tilde{\gamma}$.

Following the same MC analysis as outlined in \cref{sec:stats}, we show the $90\%$ CL limits for the case of the gradient of a scalar in \cref{fig:grad-scalar-beta}. The largest difference in this case is the increased constraining power of the Compton peak compared to the vector case shown in \cref{fig:beta-lim}. This is because of the scaling that its amplitude receives by the factor $\tilde{\gamma} > 1$.

\section{Linear Polarization Statistics}
\label{app:likelihood_linearpol}

Here we present the marginal likelihood for the linear polarization case. To get the relevant non-centrality parameter, we can set $\varphi_x = \varphi_y$ in~\cref{eq:low-t-period} to get
\begin{equation}
\hat{\lambda'}(\omega) \equiv \frac{\hat{\mathcal{P}}(\omega)}{\sigma^2} = \frac{\beta^2}{4}\Big\{\left[\hat{\alpha}_x^2 + \hat{\alpha}_y^2\right]\cos^2\phi\,(\delta_{\omega,\,s} + \delta_{\omega,\,d}) + 4 \,\hat{\alpha}_z^2 \sin^2\phi\,\delta_{\omega,\,m}\Big\}\,.
\end{equation}
Once again proceeding as in Appendices~\ref{app:deriv-marginal} and~\ref{app:likelihood_gradientscalar}, the marginal likelihood is
\begin{align}
    \mathcal{L}'' &= \int\mathrm{d}\alpha_x\mathrm{d}\alpha_y\mathrm{d}\alpha_z\mathrm{d}\,\Pi'(\alpha_x,\alpha_y,\alpha_z)\,e^{-(p+\lambda')/2}I_0(\sqrt{p\lambda'})\,,
\end{align}
which upon integrating out $\alpha$s, gives the following
\begin{equation}
    \mathcal{L}'' = \frac{2}{2A \delta_{\omega,\,m} + (2 + B \delta_{\omega,\,\Sigma})^2}\left[e^{-\frac{p}{2 + A \delta_{\omega,\,m}}} + \left(1 + \frac{p B \delta_{\omega,\,\Sigma}}{2 (2 + B \delta_{\omega,\,\Sigma})}\right) e^{-\frac{p}{2 + B \delta_{\omega,\,\Sigma}}} - e^{-\frac{p}{2}}\right]\,,
\end{equation} 
where $\delta_{\omega,\,\Sigma} \equiv \delta_{\omega,\,s} + \delta_{\omega,\,d}$, $A\equiv \beta^2 \sin^2\phi$, and $B \equiv \beta^2 \cos^2\phi / 4$. From this, it follows that the individual likelihoods are
\begin{equation}
\begin{split}
\mathcal{L}_m &= \frac{1}{2 + A} \exp{\left[-\frac{p}{2 + A}\right]}\\
\mathcal{L}_{s / d} &= \frac{2}{(2 + B)^2} \left(1 + \frac{p B}{2 (2 + B)}\right) \exp{\left[-\frac{p}{2 + B}\right]}
\end{split}
\label{eq:lik-lin}
\end{equation}

\begin{figure}[t!]
    \centering
\includegraphics{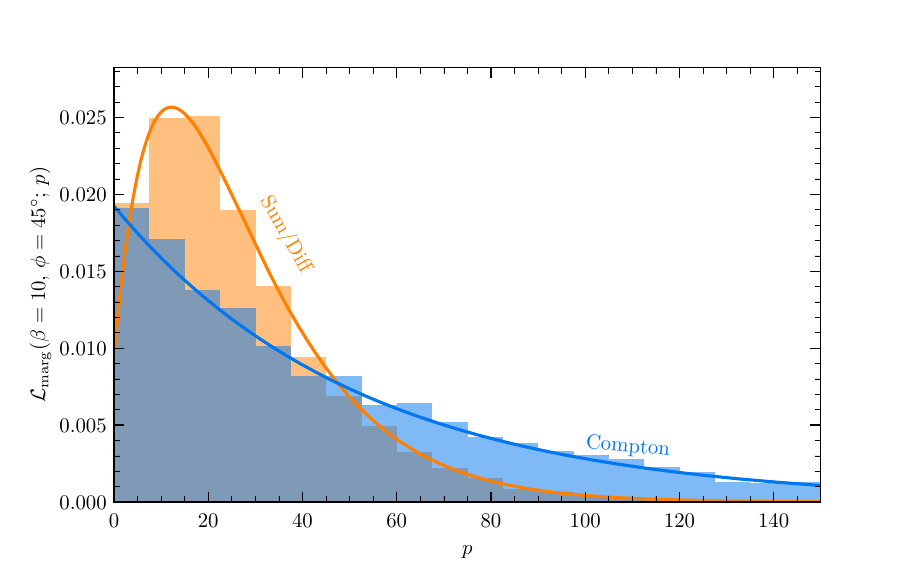}
    \caption{Example likelihoods for each of the three signal peaks once the stochastic variables have been marginalised over in the \textit{linear polarization} case. The bars show the result of a numerical simulation of the noise-normalised periodogram values beginning from \cref{eq:low-t-sig}, while the solid lines show the analytical result of \cref{eq:lik-lin}. Here, $\phi$ is the latitude of the experiment, $p$ (a random variable) is the value of the measured excess power, and $\beta$ is defined as per \cref{eq:beta}.}
    \label{fig:marginal-lik-linear}
\end{figure}

As in \cref{sec:stats}, we verify our analytical expressions for the likelihoods via a series of MC simulations. We begin from \cref{eq:low-t-sig}, this time setting all $\varphi_i$ to be equal, drawing them from a single uniform distribution, $\varphi_i \sim \mathrm{U}(0,\,2\pi)$. We draw each of the three Rayleigh variables independently from their respective distributions, as given in \cref{eq:distributions}. For each simulation, we compute the PSD as in \cref{eq:periodogram-dft} and consider the distributions of the values of each of the Compton, sum, and difference peaks, normalised by some noise level. For our simulations, we take $\mathcal{A} = 1\,[\mathcal{A}]$, $T_\mathrm{obs} = 10 T_\oplus$, $\sigma^2 = 5\,[\mathcal{A}]^2\,\mathrm{Hz^{-1}}$, and $\phi = 45^\circ$. For the purposes of fast convergence, we also take $m = 2\pi\,\si{\hertz}$ and $T_\oplus = \SI{100}{\second}$. Our results for $10^4$ simulations are shown in \cref{fig:marginal-lik-linear}, displaying excellent agreement with our derived likelihoods in \cref{eq:lik-lin}.

As in \cref{app:deriv-marginal}, we can also compute the covariance matrix for the linear polarization case. Following the approach there, we find a non-diagonal covariance matrix $\bvec{\Sigma}$ with
\begin{equation}
\bm{\Sigma} = 
\frac{\mathcal{A}^4  T_\mathrm{obs}^2}{32}
\left(
\begin{array}{@{}cc|c@{}}
    \cos^4 \phi~ & \cos^4 \phi & 0 \\
    \cos^4 \phi~ & \cos^4 \phi & 0 \\
    \hline
    0 & 0 & 8\sin^4\phi
\end{array}
\right)\,.
\label{eq:cov-linear}
\end{equation}
We thus find that the sum and difference peaks have a non-zero covariance, with the Compton peak remaining statistically uncoupled from the other two peaks. Due to this non-diagonal covariance matrix, we cannot simply write the total likelihood as in \cref{eq:lik-tot}, and we instead require a more complicated treatment accounting for the non-zero covariances.

\subsection{Effect of Elliptical versus Linear Polarization on Limits}

We comment on the effect that the linear polarization assumption has on our limits compared to the more realistic elliptical polarization treatment. Since the sum and difference peaks are correlated in the linear polarization case (c.f.~\cref{eq:cov-linear}), a full three-peak analysis is difficult to perform without accounting for the full covariance matrix. However, we can conduct a simplified, uncorrelated two-peak analysis in which we consider both the Compton peak and one of the sum/difference peaks. We can then compare the results of this analysis with a similar two-peak one done in the elliptical polarization case to learn how the limits should scale between these assumptions. Since we are only interested in this scaling, we perform a simpler Asimov analysis in which the data are assumed to be perfectly consistent with the background \cite{Cowan:2010js}. The result of an Asimov analysis is expected to asymptotically converge to the true result in the limit of high statistics.

The two-peak likelihood when considering the Compton peak and one of the two sum/difference peaks is given by
\begin{equation}
    \mathcal{L}_{\mathrm{2\text{-}peak}}^\mathrm{ell/lin}(\beta,\,\phi;\,\bvec{p}) = \funop{\mathcal{L}_m^\mathrm{ell/lin}(\beta,\,\phi;\,\bvec{p})}\funop{\mathcal{L}_{s/d}^\mathrm{ell/lin}(\beta,\,\phi;\,\bvec{p})}\,,
\end{equation}
where we have indexed the likelihoods to use for the elliptical and linear polarization cases, respectively following from \cref{eq:liks} and \cref{eq:lik-lin}. In an Asimov analysis, we replace the data vector $\bvec{p}$ with the expectation values in the background-only case, which can be shown to be $p = 2$ for each bin. In this case, using the log-likelihood-ratio test statistic given in \cref{eq:test-stat}, we will get that $\hat{\beta} = 0$ for this data. The problem then becomes finding that $\beta$ for which $q_\beta$ reaches a value that we can exclude to our desired confidence level. This is the same procedure we followed in \cref{sec:stats}, and we take $q_\beta^\mathrm{lim} \simeq 2.43$ for the $90\%$ CL limit as we found there. For comparison, we also consider the case when one wants to draw a limit at the $3\sigma$ level, equivalent to a $\sim 99.7\%$ CL limit. Solving \cref{eq:q_beta_lim}, we get that $q_\beta^\mathrm{lim} \simeq 8.49$ in this case.

\begin{figure}[t!]
    \centering
    \includegraphics{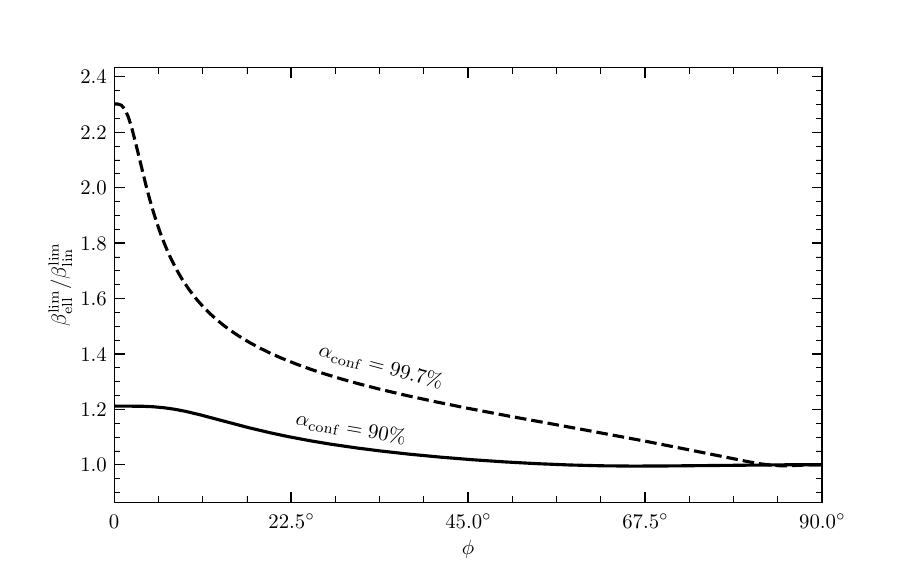}
    \caption{Scaling for limit on the dimensionless parameter $\beta$ (c.f.~\cref{eq:beta}) assuming elliptical versus linear polarization with latitude $\phi$. The $90\%$ (solid) and $99.7\%$ (dashed) limits are shown, with the latter equivalent to a $3 \sigma$ confidence level. Near the equator, the difference is more pronounced because the difference in the likelihoods is greater; the likelihoods are equal at the poles.}
    \label{fig:ell_vs_lin}
\end{figure}

We show our results in \cref{fig:ell_vs_lin}. We find that the scaling is greater towards the equator. This is because the difference in the form of the likelihoods is greater there, since only the sum/difference likelihood changes between the two polarization assumptions. At the poles, only the Compton peak is present and the two likelihoods are equivalent. This leads the two limits to converge towards the same value. At the $90\%$ CL, which we have used throughout this work, we see that the limit at worst scales by the factor $\sim 1.2$. As the CL grows, this factor increases; for example, at the $3\sigma$ level, the scaling from the linear to the elliptical assumption becomes $\sim 2.3$. Nevertheless, in the incoherent regime, we expect both the limits to match since the stochasticity of the field vanishes in that limit.
 
\section{Updated MICROSCOPE Limit}
\label{app:microscope}

We update the MICROSCOPE limit in the $B - L$ parameter space using the final results reported in \cite{MICROSCOPE:2022doy}. To the best of our knowledge, the current limit (at the time of writing) stems from \cite{AxionLimits}, which were based on the first MICROSCOPE results \cite{MICROSCOPE:2019jix}. We thank Pierre Fayet for bringing this to our attention. Following the logic outlined in their work \cite{Fayet:2017pdp,Fayet:2018cjy}, we derive the updated limit below. 

The Yukawa potential for a long-range $B - L$ force, as it is relevant for the MICROSCOPE experiment, is given by
\begin{equation}
    V_{Y,\,i}(r) = \alpha_\mathrm{EM} \varepsilon_{B - L}^2 N_\oplus N_i\,\frac{e^{-mr}}{r}\,.
\end{equation}
Here, $\alpha_\mathrm{EM} \approx 1 / 137$ is the fine-structure constant, $\varepsilon_{B - L}$ is the coupling strength of the new $B - L$ force relative to electromagnetism, $N_\oplus$ and $N_i$ are respectively the total number of neutrons in the Earth and test mass $i$, $m$ is the mass of the new gauge boson, and $r$ is the distance to the test mass from the center of the Earth. The force experienced by $i$ is then
\begin{equation}
F_{Y,\,i}(r) = -V'_{Y,\,i}(r) = \alpha_\mathrm{EM} \varepsilon_{B - L}^2 N_\oplus N_i (1 + m r)\frac{e^{-mr}}{r^2}\,.
\label{eq:yuk-force}
\end{equation}

The strength of this new force can be constrained through measurements of the Eöt-Vös parameter, defined as the normalised differential acceleration between two masses in free fall:
\begin{equation}
\eta \equiv 2 \frac{a_2 - a_1}{a_1 + a_2}\,,
\end{equation}
where $a_i$ is the acceleration of test mass $i$. Writing $a_i = (1 + \delta_i) g$ and expanding in small $\delta_i$, we get that
$$
\eta \simeq \delta_{12} \equiv \frac{a_1 - a_2}{g} =  \delta_1 - \delta_2\,.
$$
For the force given in \cref{eq:yuk-force}, we have that
\begin{equation}
    \begin{split}
        \delta_i(r) = \frac{F_{Y,\,i}(r)}{m_i g}  &= -\frac{\alpha_\mathrm{EM} M_P^2}{m_i M_\oplus}\varepsilon_{B - L}^2 N_\oplus N_i (1 + mr)e^{-mr} \\
&= -\frac{\alpha_\mathrm{EM} M_{P}^2}{u^2} \varepsilon_{B - L}^2 \left(\frac{N}{A}\right)_\oplus \left(\frac{N}{A}\right)_i (1 + mr)e^{-mr}\,,
    \end{split}
\end{equation}
where we have written the Planck mass as $M_P^2 = 1 / G_N$. We have also expressed each mass in terms of the atomic mass unit $u$ via $m_{i/\oplus} \simeq A_{i/\oplus}  u$, where $A$ is the (average) atomic mass of each body. Note that, when writing the fractional charges $N/A$, the total number of charges is irrelevant; only the average ratios over the compositions of the Earth and the test mass are relevant for the calculation. From this, we get the following expression for the Eöt-Vös parameter:
\begin{equation}
    \begin{split}
        \delta_{12} &= 
-\frac{\alpha_\mathrm{EM} M_{P}^2}{u^2} \varepsilon_{B - L}^2 \left(\frac{N}{A}\right)_\oplus \left[\left(\frac{N}{A}\right)_1 - \left(\frac{N}{A}\right)_2 \right](1 + mr)e^{-mr} \\ &\simeq -1.25 \times10^{36}\, \varepsilon_{B - L}^2 \left(\frac{N}{A}\right)_\oplus \left[\left(\frac{N}{A}\right)_1 - \left(\frac{N}{A}\right)_2 \right] (1 + mr)e^{-mr}\,.
    \end{split}
    \label{eq:delta_12}
\end{equation}

We can compare this with the value of $\delta_{12}$ measured by the MICROSCOPE collaboration to constrain $\varepsilon_{B - L}$. For two free-falling, equally massive titanium and platinum test masses, their final results yielded \cite{MICROSCOPE:2022doy}
\begin{equation}
\delta(\mathrm{Ti},\,\mathrm{Pt}) = [-1.5 \pm 2.3\,(\mathrm{stat})\pm 1.5\,(\mathrm{syst})] \times 10^{-15}\,.
\end{equation}
Adding the statistical and systematic errors in quadrature and expressing the range in $\delta$ at the $90\%$ CL, equivalent to $1.645 \sigma$, we have that
\begin{equation}
    -6.0 \times 10^{-15} \lesssim \delta(\mathrm{Ti},\,\mathrm{Pt}) \lesssim 3.0 \times 10^{-15}\,.
\label{eq:delta_12_MIC}
\end{equation}
To re-interpret this as a limit on $\varepsilon_{B - L}$ using \cref{eq:delta_12}, we must know the values of $(N / A)_\oplus$ and $(N / A)_\mathrm{Ti} - (N / A)_\mathrm{Pt}$. We retrieve both of these values from \cite{Fayet:2017pdp}, given as $0.5138$ and $-0.05625$, respectively. This gives a positive value for $\delta_{12}$, and we therefore compare it with the positive value in the range shown in \cref{eq:delta_12_MIC}. Our limit is then given by
\begin{align}
    \lvert \varepsilon_{B - L} \rvert &\lesssim 8.93 \times 10^{-19}\,  \left(\frac{N}{A}\right)_\oplus ^{-1/2} \left[\left(\frac{N}{A}\right)_\mathrm{Ti} - \left(\frac{N}{A}\right)_\mathrm{Pt} \right]^{-1/2} \frac{e^{mr / 2}}{\sqrt{1 + mr}} \nonumber\\
    &\simeq 2.89 \times 10^{-25} \frac{e^{mr / 2}}{\sqrt{1 + mr}}\,.
\end{align}
Finally, comparing the gauge coupling $g_{B - L}$ to $\varepsilon_{B -L}$ via $\lvert g_{B - L} \rvert = (4 / 5) \lvert \varepsilon_{B - L} \rvert e$ \cite{Fayet:2017pdp}, we arrive at the result
\begin{equation}
    \lvert g_{B - L} \rvert \lesssim 7.0 \times 10^{-25} \frac{e^{m r / 2}}{\sqrt{1 + m r}}\,.
\end{equation}
To evaluate the remaining expression, we use the fact that the MICROSCOPE experiment operated at a height of $710\,\mathrm{km}$. This gives $r \simeq 7066\,\mathrm{km}$, yielding
\begin{equation}
m r \simeq \frac{m}{\SI{2.79e-14}{\electronvolt}}\,.
\end{equation}
We show the final limit in \cref{fig:b-l}.

\end{document}